\documentclass{pasj00} 

\begin{document}
\SetRunningHead{Yoshiaki {\sc Sofue}}{Rotation  Curve in the Galactic Center}
\Received{2013/mm/dd}  \Accepted{2013/mm/dd} 

\def\kms{km s$^{-1}$}  \def\Msun{M_\odot} \def\deg{^\circ} \def\lv{(l, v)} \def\be{\begin{equation}} \def\ee{\end{equation}} \def\bc{\begin{center}} \def\ec{\end{center}} \def\co{ $^{12}{\rm CO}(J=1-0)$ } \def\cs{ ${\rm C}^{32}{\rm S}(J=1-0)$ } \def\dv{de Vaucouleurs }

\title{Rotation Curve and Mass Distribution in the Galactic Center --- From Black Hole to Entire Galaxy ---}
\author{Yoshiaki {\sc Sofue}$^{1,2}$ }  
\affil{
1. Institute of Astronomy, The University of Tokyo, Mitaka, 181-0015 Tokyo, \\ 
2. Department of Physics, Meisei University, Hinoshi-shi, 191-8506 Tokyo\\
Email:{\it sofue@ioa.s.u-tokyo.ac.jp}
 }


\maketitle

\begin{abstract} 
Analyzing high-resolution longitude-velocity (LV) diagrams of the Galactic Center observed with the Nobeyama 45-m telescope in the CO and CS line emissions, we obtain a central rotation curve of the Milky Way. We combine it with the data for the outer disk, and construct a logarithmic rotation curve of the entire Galaxy. The new rotation curve covers a wide range of radius from $r \sim 1$ pc to several hundred kpc without a gap of data points. It links, for the first time, the kinematical characteristics of the Galaxy from the central black hole to the bulge, disk and dark halo. Using this grand rotation curve, we calculate the radial distribution of surface mass density in the entire Galaxy, where the radius and derived mass densities vary over a dynamical range with several orders of magnitudes. We show that the galactic bulge is deconvolved into two components: the inner (core)  and main bulges. Both the two bulge components are represented  by exponential density profiles, but the de Vaucouleurs law was found to fail in representing the mass profile of the galactic bulge.
\\
{\bf Note: Preprint with full figures is available from\\
http://www.ioa.s.u-tokyo.ac.jp/$\sim$sofue/htdocs/2013rc/}

\end{abstract}

\section{Introduction}

Determination of the mass distribution in the Galaxy is one of the most fundamental subjects in galactic astronomy, and is usually obtained by analyzing rotation curves (Sofue and Rubin 2001). The rotation curve from the inner disk to the dark halo and their mass distributions have been obtained with significant accuracy (Sofue et al 2009; Sofue 2012; Honma et al. 2012; and the literature therein). The innermost mass structure within a few pc around the central black hole has been extensively studied by analyzing stellar kinematics (Crawford et al. 1985; Genzel and Townes 1987; Rieke and Rieke 1988;  Lindqvist et al. 1992; Genzel et al. 1994, 2010; Ghez et al. 2005, 2008;  Gillessen et al. 200).

The mass structure between the central black hole and the disk, and therefore, the dynamical mass structure inside the bulge, is not thoroughly studied. We derive the rotation curve in the Galactic Center, which has remained as the last unresolved problem of the rotation curve study of the Galaxy.  We derive a central rotation curve using longitude-velocity diagrams obtained by the highest resolution molecular line observations. The curve will be deconvolved into classical mass components: the black hole, bulge, disk, and dark halo.  During the analysis, we show that the \dv ($e^{-(r/a)^{1/4}}$) law cannot fit the bulge's mass structure, and that the bulge is composed of two concentric mass components with exponential density profile ($e^{-r/a}$). The fitted parameters will become the guideline to analyze perturbations often highlighted as non-circular motions and bar.

The dynamical parameters of the Galaxy to be determined from observations are summarized in table \ref{parameters} in Appendix. In the present paper we try to fix the parameters (1) to (10) in the table for the most fundamental axisymmetric part.  Non-circular motions have often been stressed in the discussion of central dynamics (Binney et al. 1991; Jenkins and Binney 1994; Athnasoula 1992; Burton and Liszt 1993). However, discussing non-axisymmetric dynamics first is akin to calculating epicyclic frequency without angular velocity. The present analysis is limited only to items (1) to (10) for the axisymmetric part, which describe the first approximation of the galactic structure. The second-order parameters (11) to (27) are beyond the scope of this paper. Although the accuracy of the obtained result may not be as good as that of the outer disk rotation curve, the present parameters will become the basis for analyses of the second-order parameters such as bars and arms.

The galactocentric distance and the circular velocity of the Sun are taken to be  $(R_0, V_0)$=(8.0 kpc, 200 \kms).   We also examine a case for the newest values of $(R_0, V_0)$=(8.0 kpc, 238 \kms) obtained by recent VERA-VLBI observations (Honma et al. 2012).
 
\section{Longitude-Velocity Diagrams}

\begin{figure}
\begin{center}  
\includegraphics[width=4cm,angle=270]{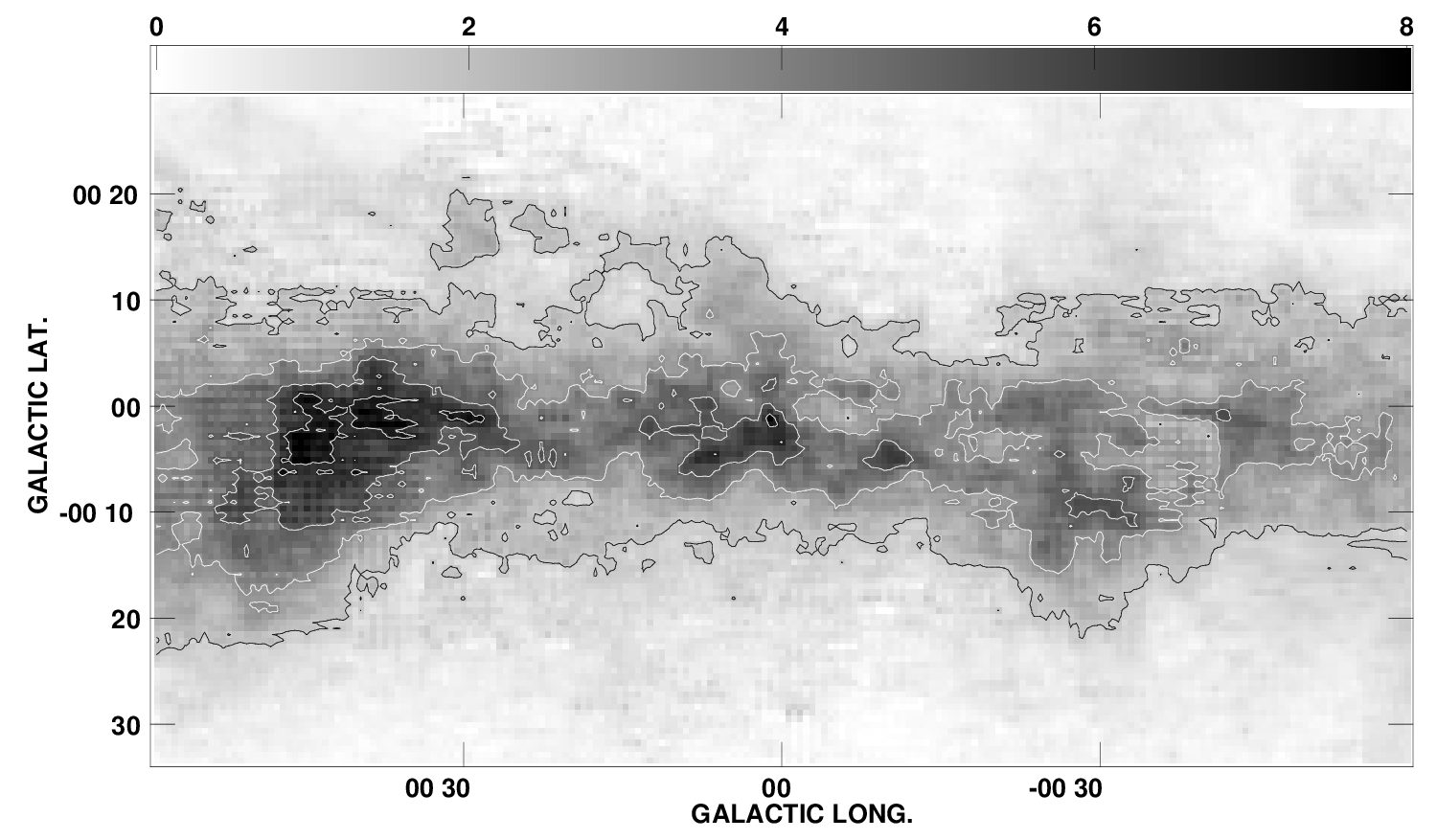}    
\ec
\caption{
Integrate intensity map of the $^{12}{\rm CO}(J=1-0)$ line (115.27 GHz) made from data by Oka et al. (1998).  
} 
\label{Ico} 

\begin{center} 
\includegraphics[width=7cm]{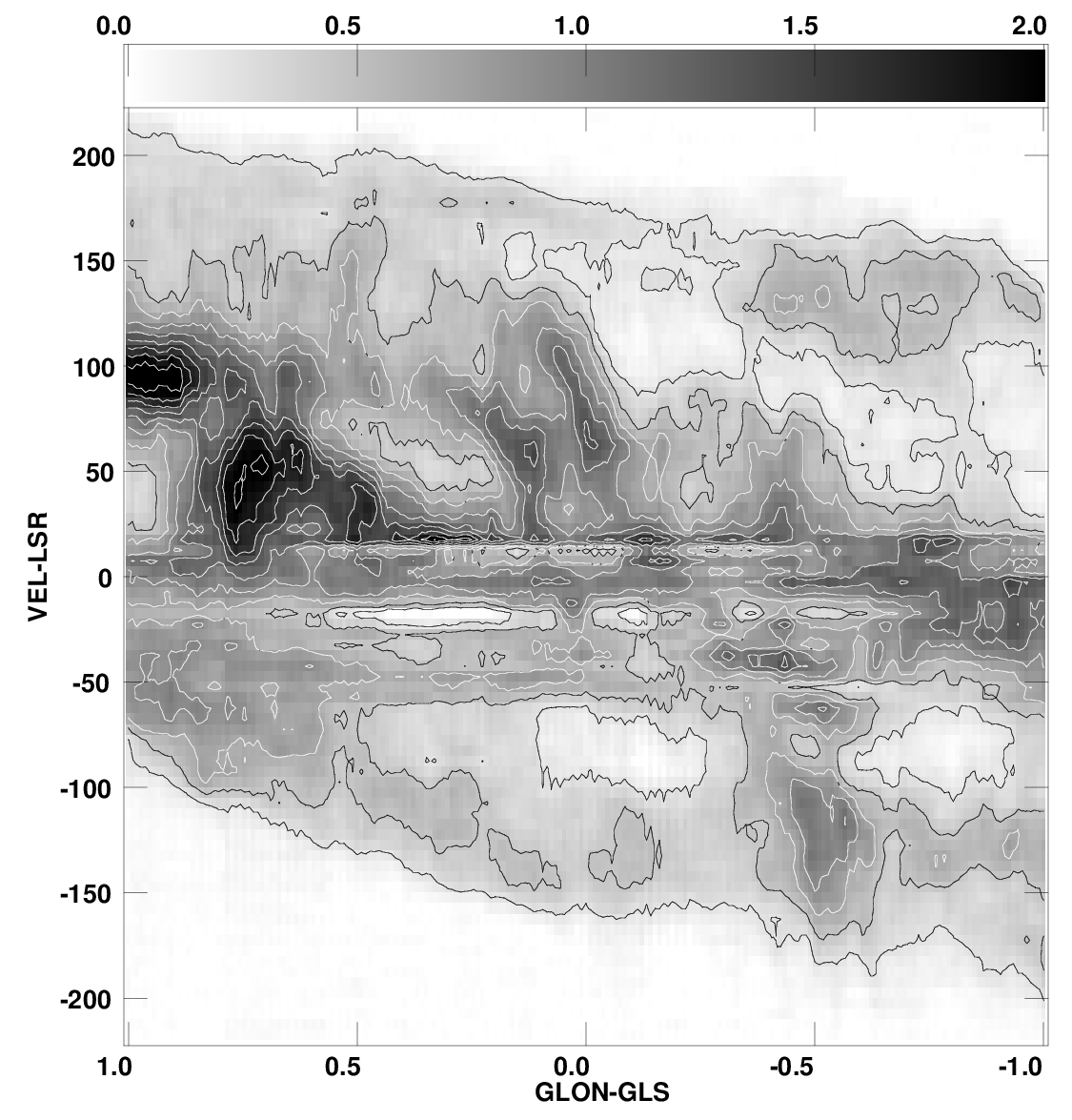}   
\end{center}
\caption{Longitude-velocity (LV) diagram in the CO line averaged over latitudes from $b=-30'$ to $+30'$ (data from Oka et al. 1998).} 
\label{LVav}
\end{figure}

Figure \ref{Ico} shows the intensity distribution of molecular gas in the \co line (115.27 GHz) for the central $\pm 1^\circ$ region of the Galactic Center as produced from the survey data using the Nobeyama 45-m telescope by Oka et al. (1998). Although the telescope beam was $15''$, the observations were obtained with $34''$ gridding, resulting in an effective resolution of $37''$. 

Figure \ref{LVav} shows a longitude-velocity (LV) diagram of the central molecular disk averaged from $b=-20'$ to $+20'$. The LV diagram is characterized by two major structures. The tilted LV ridges are the most prominent features, which correspond to dense nuclear gas  arms in the disk (Sofue 1995a).  The tilted LV ridges shares most of the intensities, while the high-velocity components shares only a few percents of the total gas (Sofue 1995a,b). The tilted LV ridges correspond to the major disk at low latitudes making ring-like arms. The high-velocity arcs come from fainter structures extending perpendicular to the disk at higher latitudes.

The high-velocity arcs are visible as parts of an ellipse corresponding either to parallelogram (Binney et al. 1991) or to an expanding molecular ring (Kaifu et al. 1972; Scoville 1972; Sofue 1995b). The LV ridges and the high-velocity arcs must not be physically related to each other, but are present at different locations, since their distributions and kinematics cannot be originating from the same gas cloud, unless different gas streams can cross each other. In the present analysis, we focus on the LV ridges corresponding to the major disk.

\begin{figure}
\bc
\includegraphics[width=8cm]{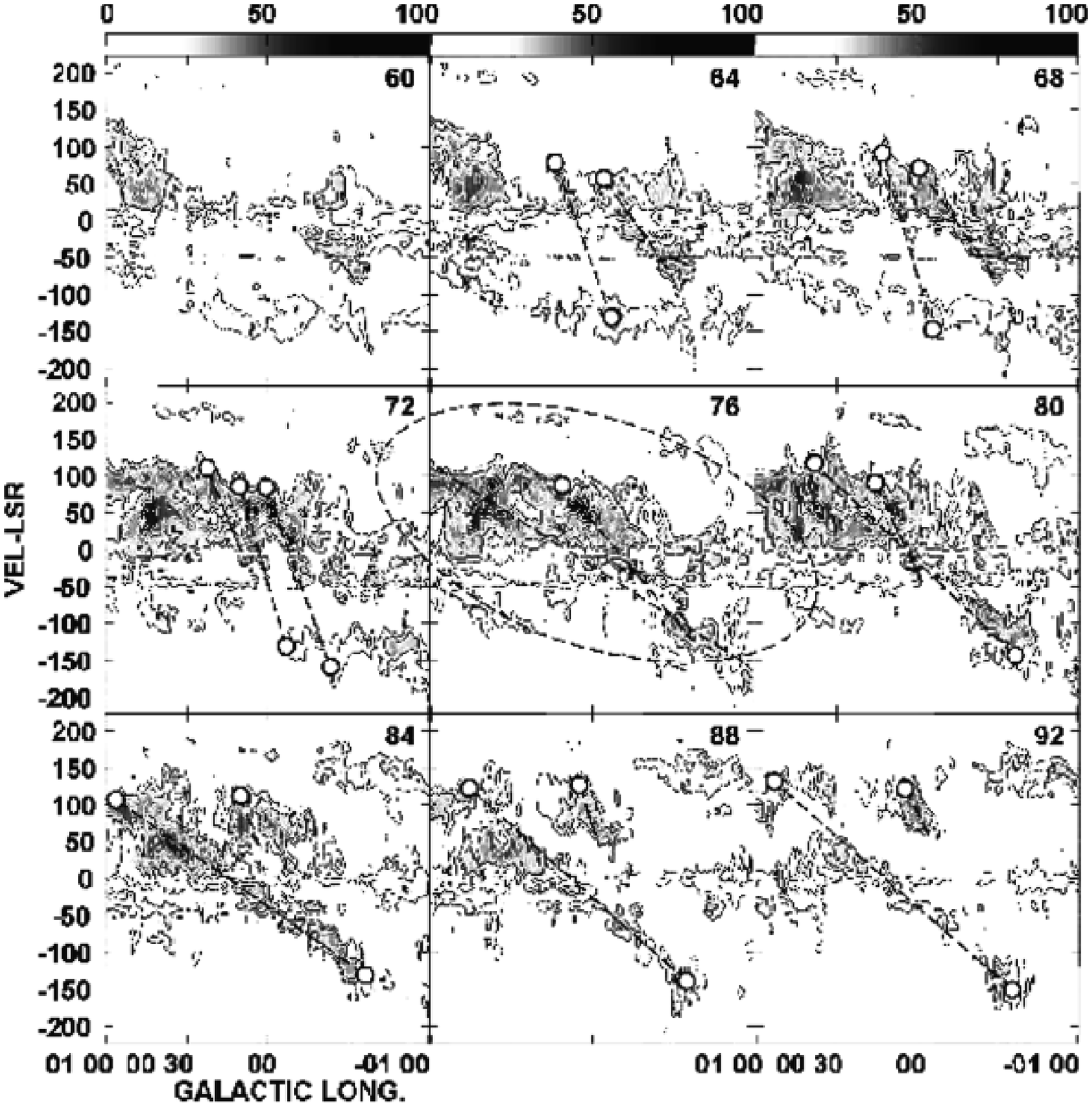}   
\end{center}
\caption{Longitude-velocity diagrams at different latitudes in the \co line (115.27 GHz; Oka et al. 1998) from $b=-11'.9$ (top-left) to $b=+6'.23$ (bottom-right) at $2'.26$ latitude interval. Overlaid are traced LV ridges and corresponding terminal velocity tracing.} 
\label{LVco}  

\bc
\includegraphics[width=8cm]{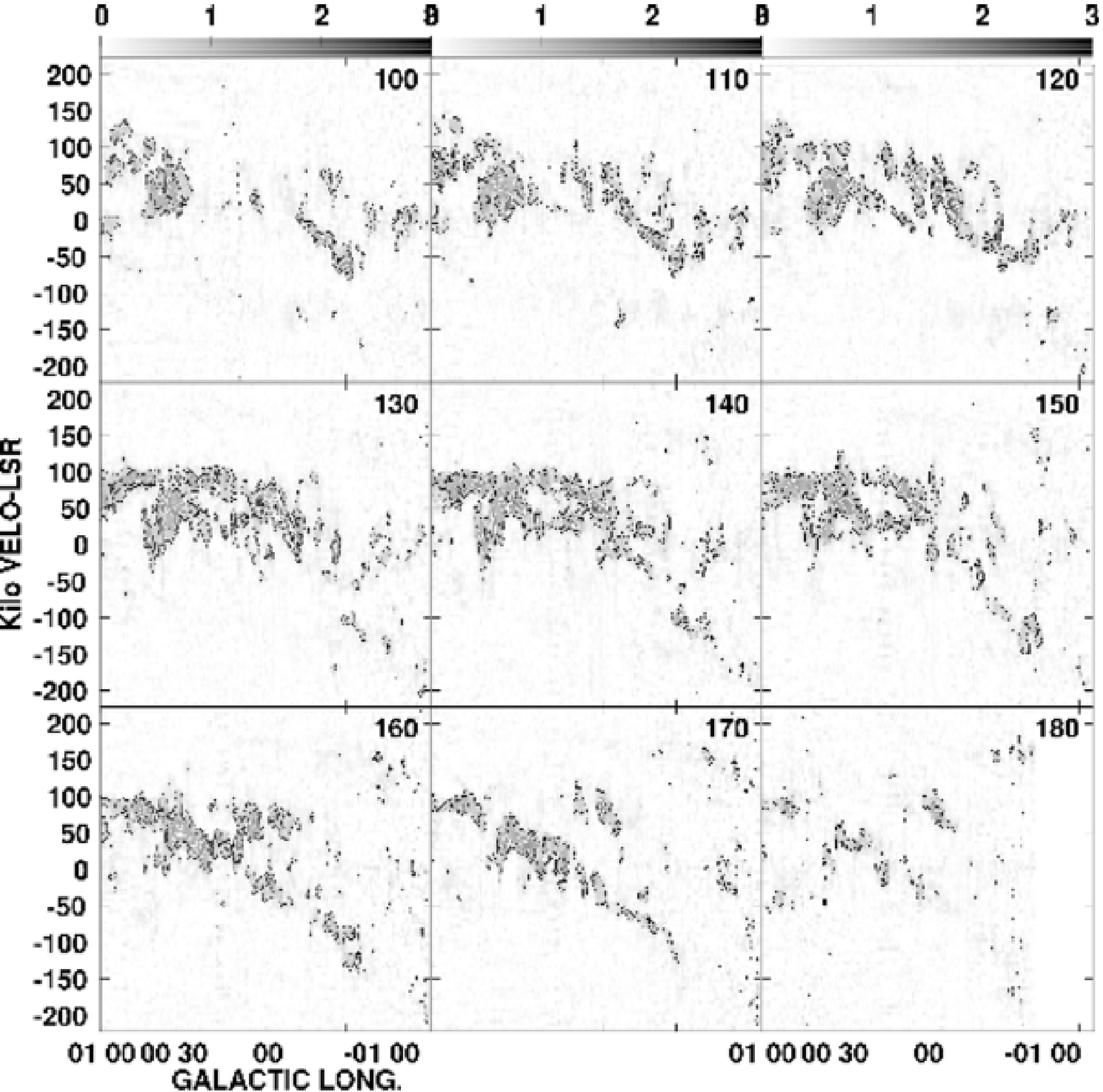}   
\ec
\caption{Same as figure \ref{LVco} but in \cs line (48.99 GHz) data  from Tsuboi et al. 1999), showing denser gas kinematics, with latitude interval $1'.553$ from $b=-9'.033$ (top left) to $+4'.187$ (bottom right).} 
\label{LVcs}
\end{figure}

In order to examine the kinematics of the molecular disk in more details, and to see if the major gas disk is characterized by the tilted-ridge structures, we present LV diagrams at different latitudes by slicing the disk. Figures \ref{LVco} and \ref{LVcs} show the LV diagrams in the \co (115.27 GHz; Oka et al. 1998) and \cs  (48.99 GHz; Tsuboi et al. 1999) line emissions. The CS data cube had an effective resolution of $46''$ arising from the telescope beam of $35''$ of the 45-m telescope and the grid spacing of $30''$  during the observations.

The figures show that most of the molecular gas is distributed on the tilted ridges, representing the central molecular zone (CMZ), or the main Galactic Center disk. The major LV ridge features  are indicated in the figures such as the ridge of dense molecular gas running from $\lv=(-0.6, -150)$ to $(0.2, +130)$ (in degree and \kms). 

The tilted ridges make the fundamental structure in the LV diagram, whereas the parallelogram/expanding shell is much fainter. From comparison of figures \ref{LVco} and \ref{LVcs}, we learn that the denser gas represented by CS line is more strongly concentrated in the tilted ridges. Also, the high-velocity arcs are hardly seen in the CS line emission. We consider that the dynamics of the Galactic Center disk is better represented by these dense molecular features than by the high-velocity arcs. In this paper, we thus analyze the tilted LV molecular gas ridges. 

\begin{figure}
\begin{center}   
\includegraphics[width=6.5cm]{LVcohich.ps}   
\includegraphics[width=6cm]{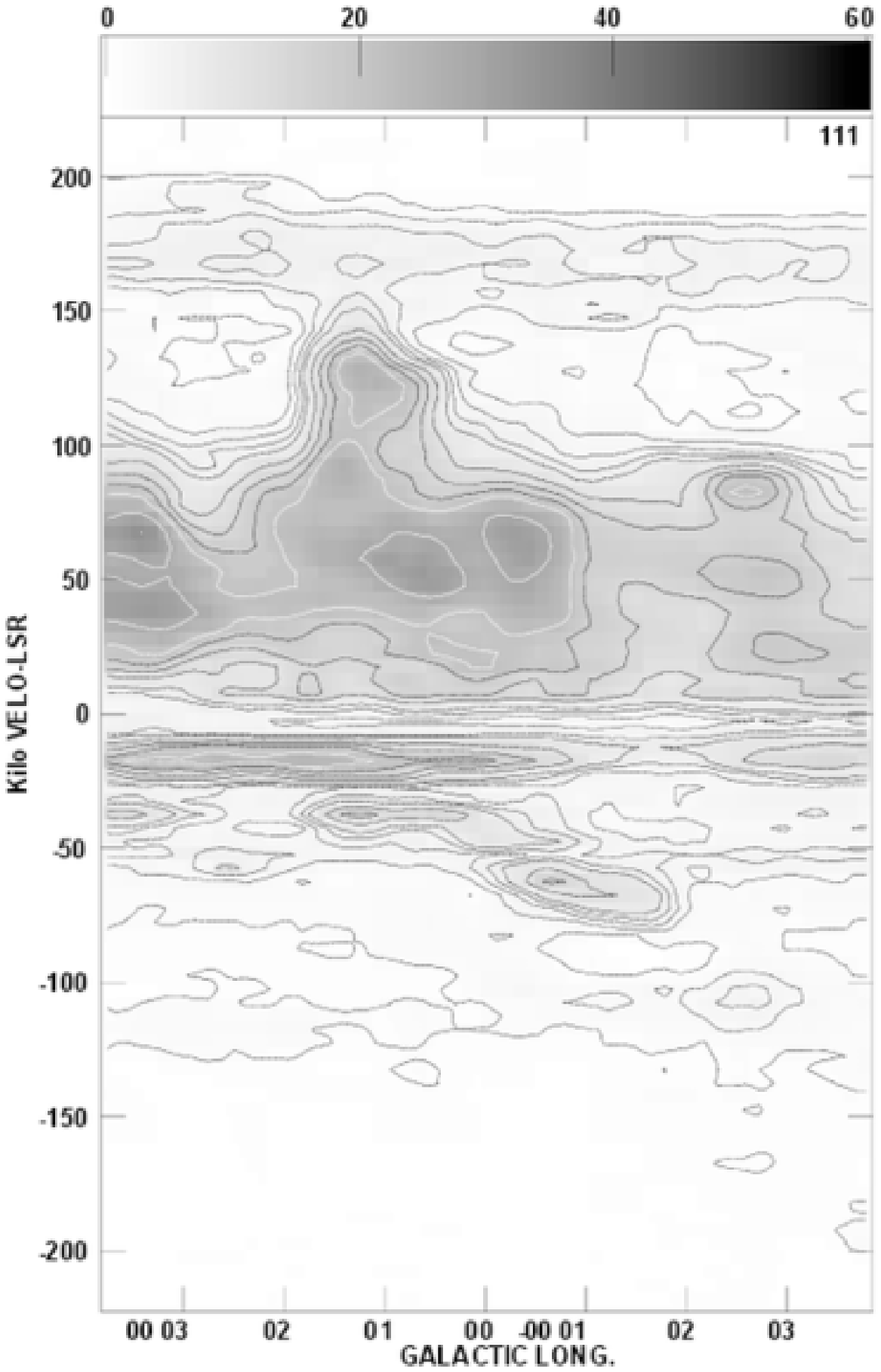}
\end{center} 
\caption{Top: High-resolution LV diagrams of the Galactic Center in the $^{12}$CO ($J=1-0$) line  observed with the NRO 45-m telescope at resolution of $15''$ at latitude interval $19''.56$  from $b=3'14''$ (top left) to $0'37''$ (bottom right).
Bottom: Same for the most clear LV ridge at $b=-2'08''$. } 
\label{LVhigh}
\end{figure}

Figure \ref{LVhigh} shows an LV diagram in the \co line emission observed using the NRO 45-m telescope at a $15''$ with $7''.5$ Nyquist-sampling gridding. In this highest resolution LV diagram, we can also recognize a tilted LV ridge running from $\lv=(-1.5', -80)$ to $(1.2', 100)$.

\begin{figure}
\bc
\includegraphics[width=5cm]{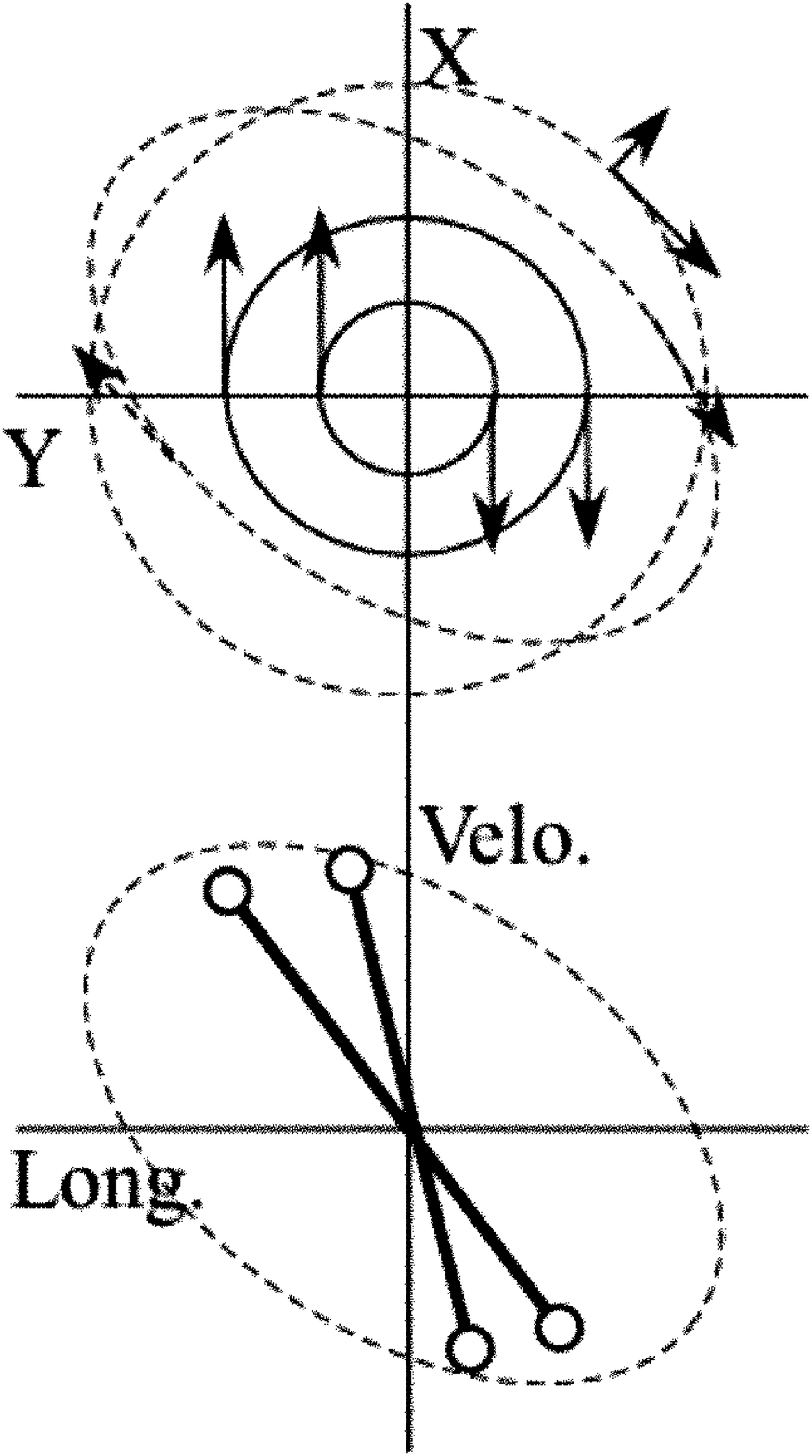}   
\ec
\caption{Circular rotation (full lines) vs streaming motion and expanding ring (dashed lines) projected on the galactic plane and LV plane.} 
\label{Oval}
\end{figure} 

\section{Rotation Curve}

\subsection{Terminal Velocities by LV Ridge Tracing}

Tilted ridges in the LV diagrams are naturally interpreted as due to arms and rings rotating around the GC, as illustrated in figure \ref{Oval}. In fact, the major LV ridges have been shown to represent ring/arm like structures in the main disk (Sofue 1995a, b). We now trace the LV ridges on the observed LV diagrams shown in figures \ref{LVco}, \ref{LVcs} and \ref{LVhigh}. By applying the terminal velocity method (Rubin and Sofue 2001) to each LV ridge, we determine rotational velocities on individual LV ridges. Here, we adopted the terminal velocity as the velocity showing the steepest gradient.  
 
 Since the original LV diagrams in the Galactic Center are superposed by various kinematical features such as molecular clouds, star forming regions (Sgr B, C, etc..), high-velocity wings (e.g. Oka et al. 1998), and foreground absorption, it was not practical to write a program to determine the terminal velocity automatically from machine-readable data. So, we read the velocities by eye, judging individually the LV behaviors in velocity and longitudinal extents. The thus measured velocities and positions included errors of the order of $\sim \pm 10$ \kms in velocity and about twice the effective angular resolutions.

The obtained terminal velocities are shown in figure \ref{GCrc}. The velocities are scattered locally by 20-30 \kms. The east-west asymmetry in velocity distribution is greater than the local scatter, and amounts to almost 30-40 \kms. Open circles in figure \ref{GCrc} shows a rotation curve produced from the thus measured terminal velocities. The filled circles in the bottom panel in figure \ref{GCrc} show running-averaged values every 1.3 times the neighboring radius with a Gaussian weighting width of 0.3 times the radius.

Standard deviations among the neighboring data points during the running-averaging process are indicated by the horizontal (position) and vertical (velocity) error bars. Although each velocity includes small error of order $\sim 10$ \kms, the running-averaged values have errors of $\sim \pm 20-30$\%. This large scatter (error) cannot be removed, because it arises from the dynamical property of molecular gas in the Galactic Center.

\subsection{Grand rotation curve from the black hole to halo}
 
\begin{figure}
\bc
\includegraphics[width=8.6cm]{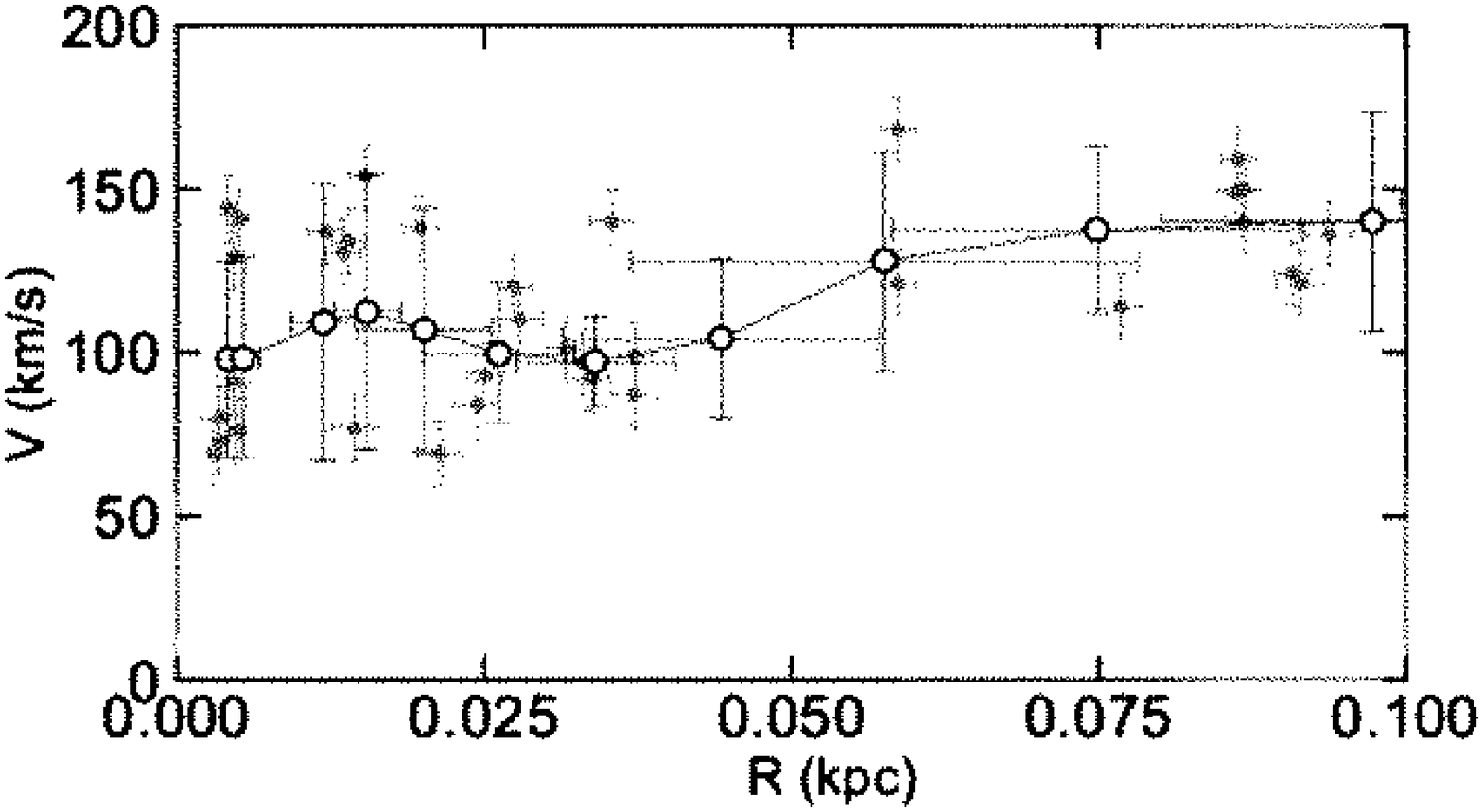}   
\vskip -5mm
\includegraphics[width=9cm]{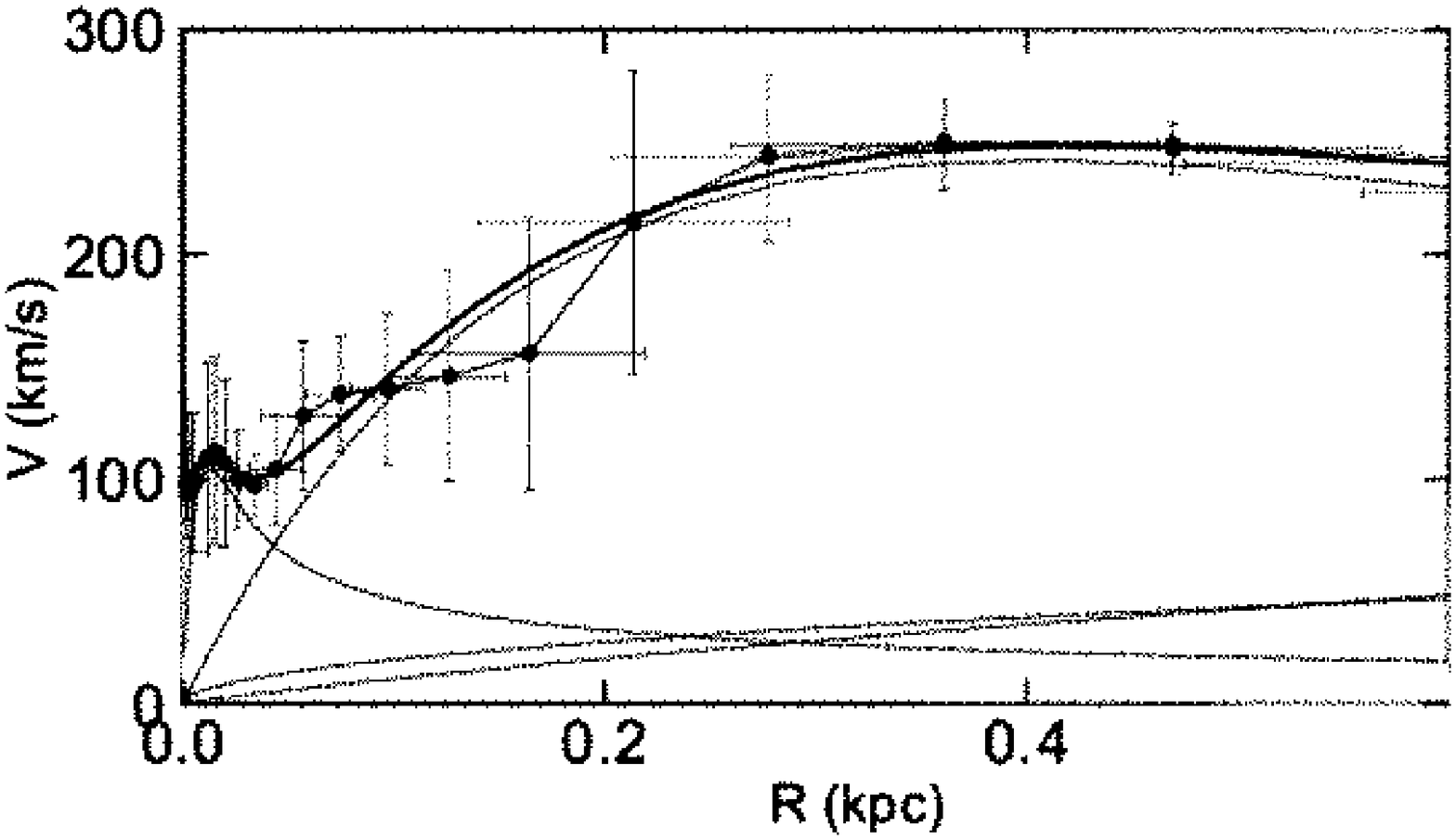}  
\ec
\caption{Top: Rotation velocities (grey dots) in the Galactic Center obtained by LV ridge terminal method. The error bars represent effective spatial resolutions taken to be twice the effective angular resolution in the data ($15''$ for our innermost CO data by 45-m telescope; $37''$ for CO data by Oka et al. (1989); $48''$ for CS data by Tsuboi et al. (1991)), and eye-estimated terminal velocity errors of $\pm 10$ \kms. Open circles show running-averaged values every 3 points using the neighboring 5 data points. The error bars denote standard deviation in the averaged data used in each plotted point. 

Bottom: Same up to 0.6 kpc, but Gaussian-running averaged velocities combined with the data by Sofue et al. (2009). Deconvolved components (inner and main bulges, disk and dark halo) are indicated by the full lines.} 
\label{GCrc}
\end{figure}

\begin{figure} 
\bc
\includegraphics[width=9cm]{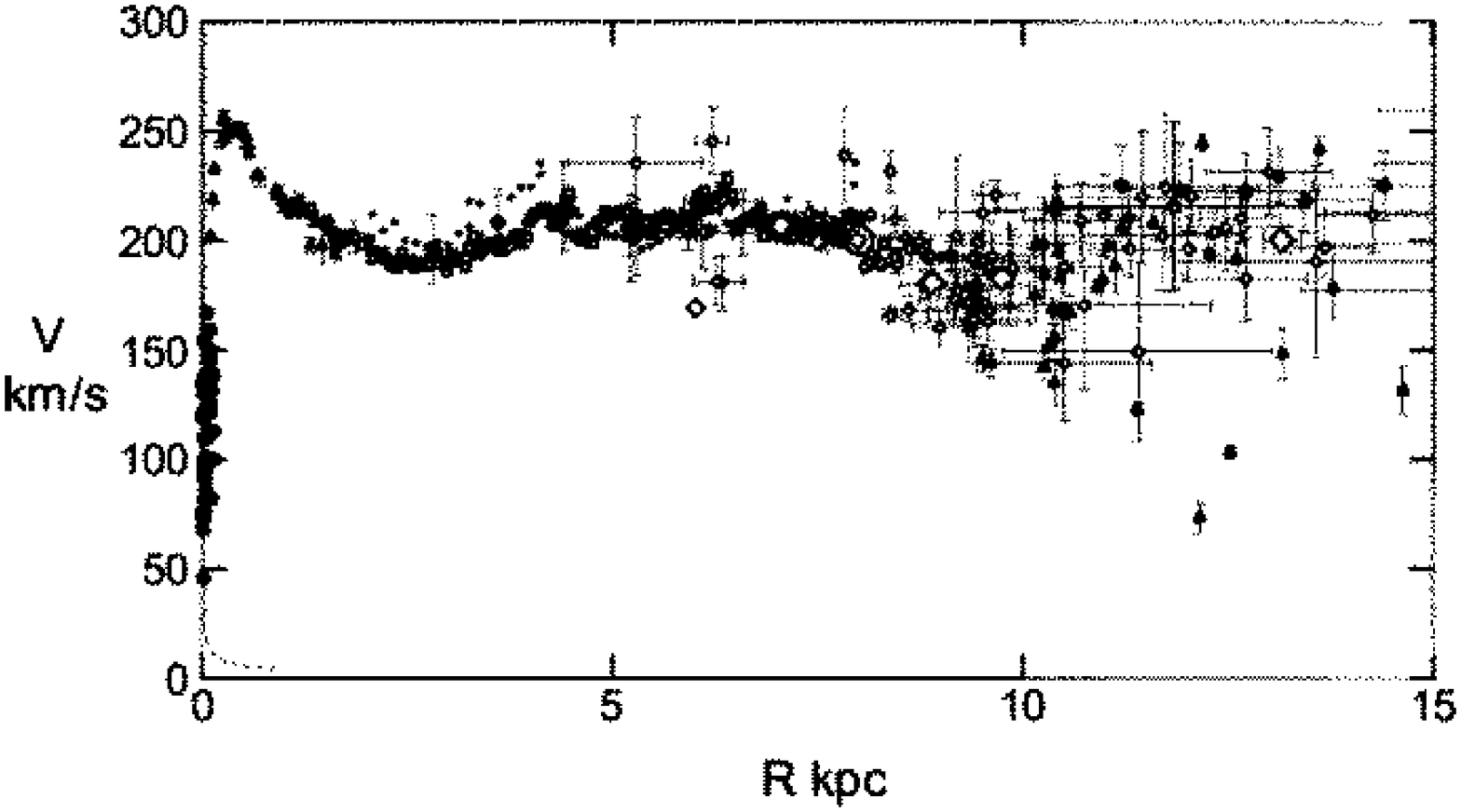}   
\includegraphics[width=9cm]{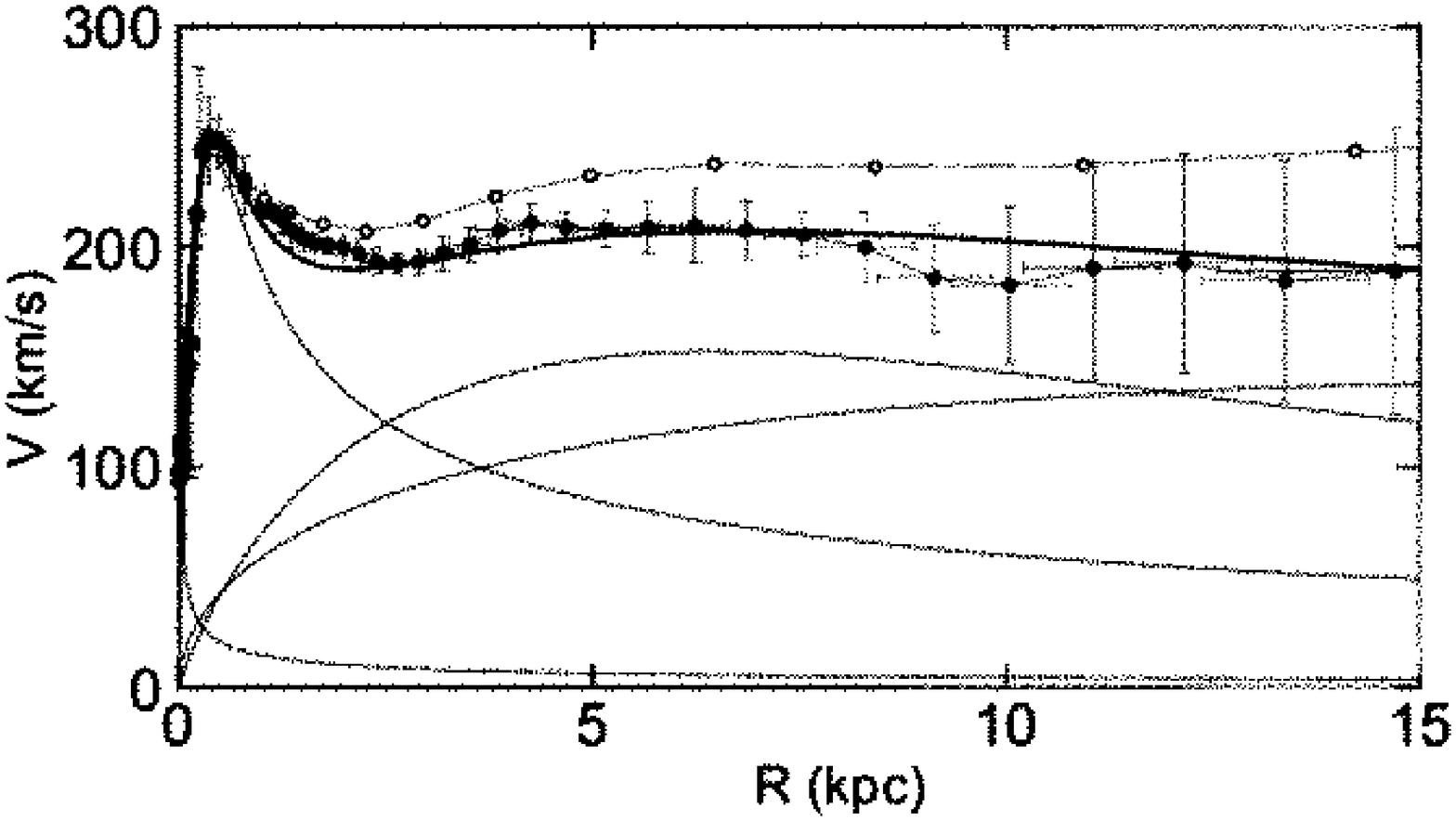}  
\ec 
\caption{Top: Rotation curve of the Galaxy from Sofue et al. (2009) combined with the running-averaged rotation velocities in the Galactic Center. Bottom: Same, but smoothed curve by Gaussian-running mean. Deconvolved components are indicated by the full lines. The Upper curve with open circles is a smoothed rotation curve for the newest values of ($R_0=8.0$ kpc, $ V_0=238$ \kms).} 
\label{RC}
\end{figure}

Figure \ref{RC} shows the obtained running-averaged rotation curve for the central region combined with the rotation curve of the whole galactic disk. The plotted data beyond the bulge were taken from our earlier papers (Sofue et al 2009; Sofue 2009). The bulge component with its peak at $R\sim 0.3$ kpc seems to decline toward the center faster than that expected for the de Vaucouleurs law as calculated by Sofue (2012). Figure \ref{GCrc} is the same within 0.5 kpc, which shows that the velocity is followed by a flat part at $R\sim 0.1$ to 0.01 kpc. The rotation velocity within the bulge at $R\le \sim 0.5$ kpc seems to be composed of two separate components, one peaking at $R\sim 0.3$ kpc, and the other a flat part at $R\sim 0.01-0.1$ kpc. 

\begin{figure}
\bc 
\includegraphics[width=9cm]{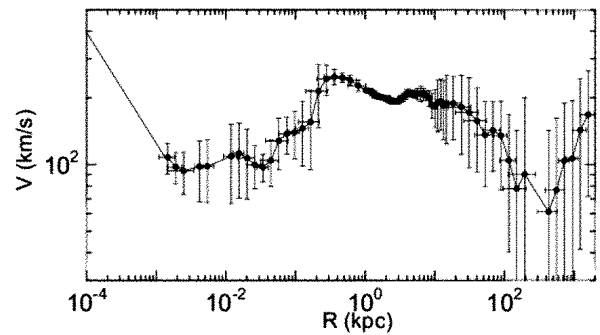}    
\ec
\caption{Logarithmic rotation curve of the entire Galaxy by combining the present data with the outer curve from Sofue et al. (2009). The thin line represents the central black hole of mass $3.6\times 10^6 \Msun$. The innermost three dots are interpolated values using the data and a Keplerian curve for the black hole. }  
\label{LogRC} 
\end{figure}

Figure \ref{LogRC} is a logarithmic plot of the measured rotation velocities combined with the grand rotation curve of the Galaxy covering the dark halo (Sofue 2012). The logarithmic representation is essential to analyze the central part, as it enlarges the radial scale toward the center according to the variation of  dynamical scale. In the figure, the disk to bulge rotation data have been adopted from the existing HI and molecular line observations (Burton and Gordon 1978; Clemens 1985; and the literature in Sofue (2009)). The inner straight dashed line represents the central massive black hole of mass $3.6 \times 10^6 \Msun$ (Genzel et al. 2000; Ghez et al. 2005; Gillesen et al. 2009).

Here, in figure \ref{LogRC}, the observed rotation velocities have been running averaged by Gaussian convolution around each representative radius at every $1+\epsilon$ times the neighboring inner point with a Gaussian width of $\pm \eta$ times the radius. Here we take $\epsilon=\eta=0.1$ for radius $3<R<15$ kpc where data points are dense, and otherwise 0.3.
For an initial radius $a$, the $j$-th radius is given by
\be
r=(1+\epsilon)^{(j-1)} a,
\ee
and the Gaussian width for running average (Gaussian convolution) is taken as
\be
\Delta r=\eta r.
\ee
The mean value of an observable $f(r)$, which is either the radius $r$ or the velocity $V$, at $r$ is calculated as
\be
f(r)={\Sigma f_i w_i \over {\Sigma w_i}},
\ee
where $w_i$ is the weight given by
\be
w_i={\rm exp}\left[-\left({{r_i-r} \over {\Delta r}}\right)^2\right].
\ee
The statistical error of the observable is calculated by
\be
\sigma_i=\left[ {\{\Sigma (f_i-f)^2 w_i \}\over {\Sigma w_i -1}}\right]^{1/2} .
\ee
The curve is drawn to connect the central rotation curve smoothly to the Keplerian law by the central massive black hole. This figure demonstrates, for the first time, continuous variation of the rotational velocity from the central black hole to the dark halo. Table \ref{rctab} lists the obtained rotation velocities.

\section{Deconvolution of Rotation Curve}
 
\subsection{Rotation Curve around the Galactic Center} 

The logarithmic rotation curve shows the central disk behavior more appropriately than the linear rotation curves. The main bulge component has a velocity peak at $r\simeq 400 pc$ with $V=250$ \kms. It declines toward the center steeply, followed by a plateau-like hump at $r \sim 30-3$ pc. The plateau-like hump is then merged by the Keplerian rotation curve corresponding to the central black hole at $r< \sim 2$ pc. Here we use a black hole mass of $M_{\rm BH}=4\times 10^6\Msun$, taking the mean of the recent values converted to the case for  $R_0=8.0$ kpc, i.e., $2.6-4.4 \times 10^6\Msun$ (Genzel et al. 2000, 2010),  $4.1-4.3\times 10^6\Msun$ (Ghez et al. 2005), and $3.95 \times 10^6\Msun$ (Gillessen et al. 2009).
 
\subsection{Broad velocity maximum by \dv law}

\begin{figure*}
\bc
\includegraphics[width=12cm]{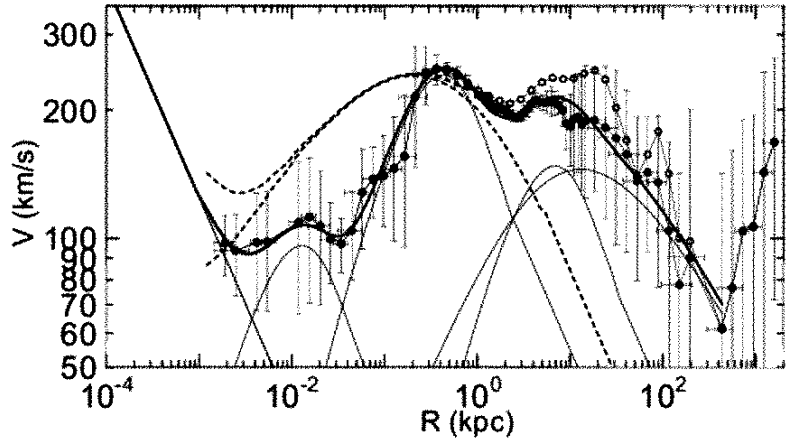}   
\ec
\caption{Logarithmic rotation curve of the Galaxy compared with model curves and deconvolution into mass components. Solid lines represent the best-fit curve with two exponential-spherical bulges, exponential flat disk, and NFW dark halo. The classical de Vaucouleurs bulge is shown by dashed line, which is significantly displaced from the observation. Open circles are a new rotation curve adopting the recently determined circular velocity of the Sun, $V_0=238$ \kms (Honma et al. 2012).
}
\label{LogLogComp}
\end{figure*}

We first compare the observations with the best-fit \dv law for the galactic bulge as obtained in our previous works (Sofue et al. 2009), which is shown by the dashed lines in figure \ref{LogLogComp}. It is obvious that the model fit is not sufficient in the inner several hundred pc.
It is valuable to revisit \dv rotation curve, which is represented by $\Sigma \propto e^{-(r/a)^{1/4}}$ with $\Sigma$ and $a$ being the surface mass density and scale radius. By definition scale radius $a$ used here is equal to $R_{\rm b}/3460.$, where $R_{\rm b}$ is the half-surface mass radius used in usual \dv law expression (e.g. Sofue et al. 2009).

Since $\Sigma$ is nearly constant at $r\ll a$, the volume density varies as $\propto 1/r$ and the enclosed mass $\propto r^2$. This leads to circular velocity $V =\sqrt{GM/r} \propto r^{1/2}$ near the center. Thus, the rotation velocity rises very steeply with infinite gradient at the center. It should be compared with the mildly rising velocity as $V \propto r$ at the center in the other models. 

At large $r >a$, the \dv law has slower density decrease due to the weaker dependence on $r$ ($r^{1/4}$ effect) than the other models. This leads to more gentle decrease after the maximum. Figure \ref{Dfunc} shows normalized behaviors of rotation velocity for \dv and other models. 
As the result of steeper rise near the center and slower decrease at large radii, the \dv rotation curve shows a much broader maximum in logarithmic plot compared to the other models. 

We here define the half-maximum logarithmic velocity width by
$
\Delta_{\rm log}={\rm log} ~r_2-{\rm log}~r_1,
$
where $r_2$ and $r_1$ ($r_2>r_1$) are the radii at which the rotational velocity becomes equal to a half of the maximum velocity. From the calculated curves in figure \ref{Dfunc}, we obtain $\Delta_{\rm log} = 3.0$ for \dv, while $\Delta_{\rm log}=1.5$ for the other models. Thus the \dv's logarithmic curve width is twice the others, and the curve's shape is much milder. Note that the logarithmic curve shape keeps the similarity against changed parameters such as the mass and scale radius.

In figure \ref{LogLogComp} (top) we show LogRC calculated for the \dv model by dashed lines and compare with the observations. It is obvious that the \dv law cannot reproduce the observations inside $\sim 200$ pc. Note that the shape of the curve is scaling free in the logarithmic plot. The \dv curve can be shifted in both directions by changing the total mass and scale radius, but the shape is kept same.

\subsection{Exponential spheroid model}

Since the \dv law was found to fail to fit the observed LogRC, we now try to represent the inner rotation curve inside $\sim 100$ pc by different models. We propose a new functional form for the central spheroidal component, which we call the {\it exponential sphere} model. In this model, the volume mass density $\rho$ is represented by an exponential function of radius $r$ with a scale radius $a$ as
\be
\rho(r)=\rho_{\rm c} e^{-r/a}.
\ee
The mass involved within radius $r$ is given by
\be
M(R)=M_0 F(x),
\ee
where $x=r/a$ and
\be
F(x)=1-e^{-x}(1+x+x^2/2).
\ee
The total mass is given by
\be
M_0=\int_0^\infty 4 \pi r^2 \rho dr=8 \pi a^3 \rho_{\rm c}.
\ee
The circular rotation velocity is then calculated by
\be
V(r)=\sqrt{G M/r}=\sqrt{{G M_0\over a} F\left({r\over a}\right)}
\ee
where $G$ is the gravitational constant.

\begin{figure}
\bc
\includegraphics[width=9cm]{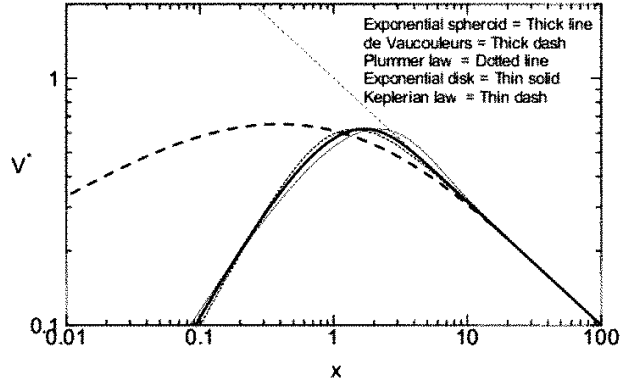}  
\ec
\caption{Comparison of normalized rotation curves for the exponential spheroid, de Vaucouleurs spheroid, and other typical models, for a fixed total mass. The exponential spheroid model is almost identical to that for Plummer law model.}
\label{Dfunc}
\end{figure}

This model is simpler than the canonical bulge models such as the de Vaucouleurs profiles. Since the density decreases faster, the rotational velocity has narrower peak near the characteristic radius in logarithmic plot as shown in figure \ref{Dfunc}. The exponential-sphere model is nearly identical to that for the Plummer's law, and the rotation curves have almost identical profiles. Hence, the results in the present paper may not be much changed, even if we adopt the Plummer potential. 


\section{Mass Distribution}

\subsection{Mass Component Fitting} 

In order to fit the observed rotation curve by models, we assume the following components, each of which should be determined of the parameters as listed in table \ref{parameters} :
\begin{itemize}
\item The central black hole with mass $M_{\rm BH}=4\times 10^6 \Msun$.
\item An innermost spheroidal component with the exponential-sphere density profile, or a central massive core.
\item A spheroidal bulge with the exponential-sphere density profile.
\item An exponential flat disk.
\item A dark halo with NFW profile.
\end{itemize}
The approximate parameters for the disk and dark halo are adopted from the current study such as by Sofue (2012), and were adjusted here in order to better fit the data. The inner two spheroidal components were fitted to the data by trial and error by changing the parameter values. 

After a number of trials, we obtained the best-fit parameters as listed in table \ref{param}. Figure \ref{LogLogComp} shows the calculated rotation curve for these parameters. The result satisfactorily represents the entire rotation curve from the central black hole to the outer dark halo.

\begin{table*}
\caption{Parameters for exponential sphere model of the bulge.}
\begin{center} 
\begin{tabular}{llll}
\hline
Mass component&Total mass ($\Msun$) & Scale radius (kpc)& Center density $(\Msun{\rm pc}^{-3})$ \\ 
\hline
Black hole & 4E6 & ---&--- \\
Inner bulge (core) & 5.0E7  & 0.0038 &3.6E4 \\
Main Bulge &    8.4E9 & 0.12 &1.9E2\\
Disk & 4.4E10 & 3.0 &15\\
Dark halo &5E10 ($r\le h$)  &$h=12.0$ & $\rho_0=0.011 $\\
\hline 
\end{tabular}
\end{center}
\label{param}
\end{table*}

We find that the fitting is fairly good in the Galactic Center, and the inner two peaks of rotation curve at $r\sim 0.01$ kpc and $\sim 0.5$ kpc are well reproduced by the two exponential spheroids. The figure also demonstrates that the present model is better than the \dv model shown by the dashed line. 
Figure \ref{RC} shows the usual presentation of the rotation curve up to 15 kpc in linear scales. The bottom panel enlarges the central several hundred pc region. 

Table \ref{param} lists the fitted  parameters for individual components. The disk and halo parameters are about the same as those determined in our earlier paper (Sofue 2012). The classical bulge is composed of two superposed components. The inner bulge, or a massive core, has a mass of $5 \times 10^7 \Msun$, scale radius of $3.8$ pc, and the central density of  $4 \times 10^4 \Msun {\rm pc}^{-3}$. The main bulge has a mass $10^{10}\Msun$, scale radius 120 pc, and central density $200 \Msun {\rm pc}^{-3}$. The central volume densities are consistent with the surface mass density (SMD) of the order of $ \sim 10^5 \Msun {\rm pc}^{-2}$ at $r\sim 3-10$ pc directly calculated from the rotation curve (Paper I). 

\subsection{Volume density}

\begin{figure*}
\bc
\includegraphics[width=8.5cm]{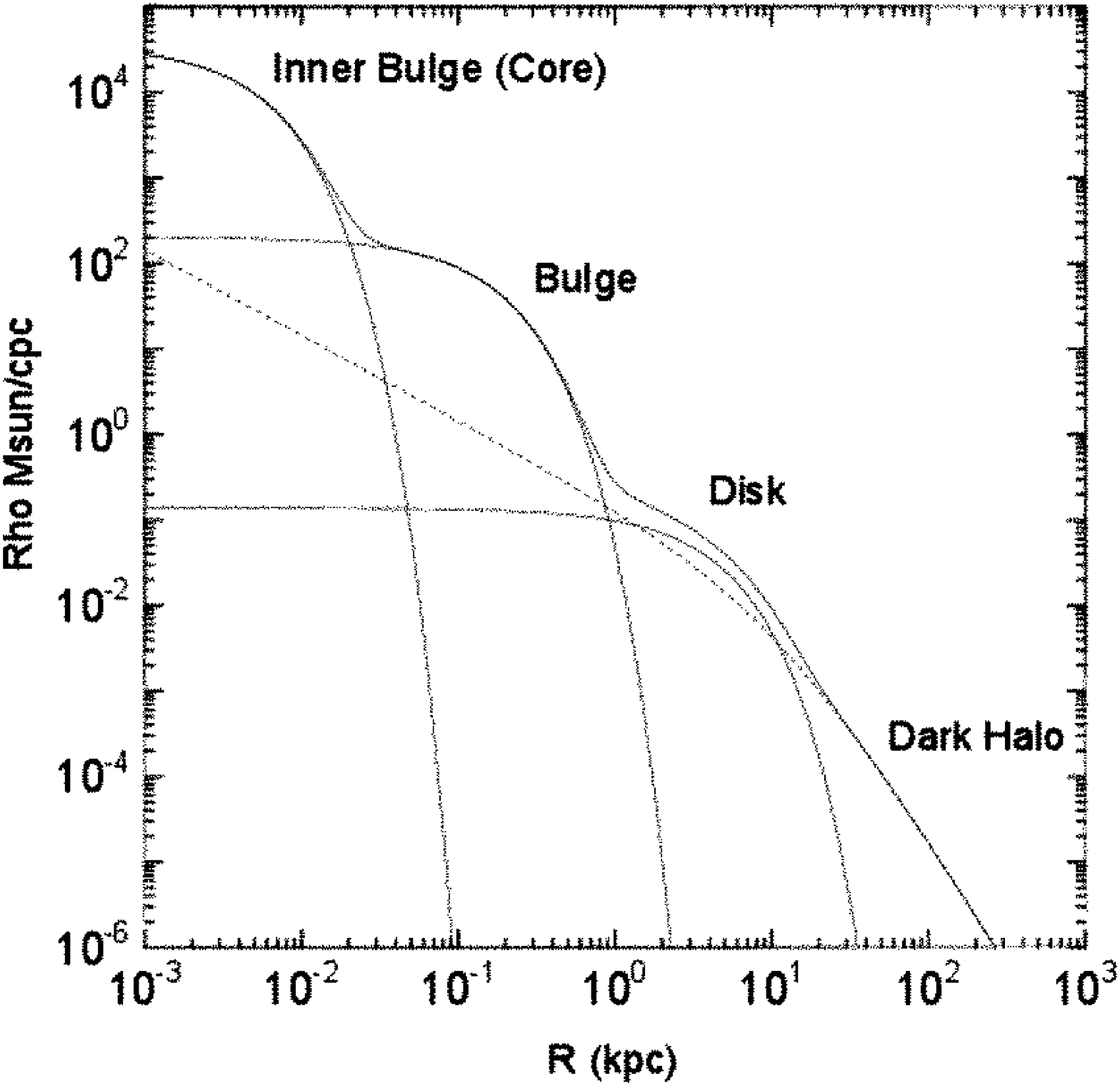}     
\includegraphics[width=8.5cm]{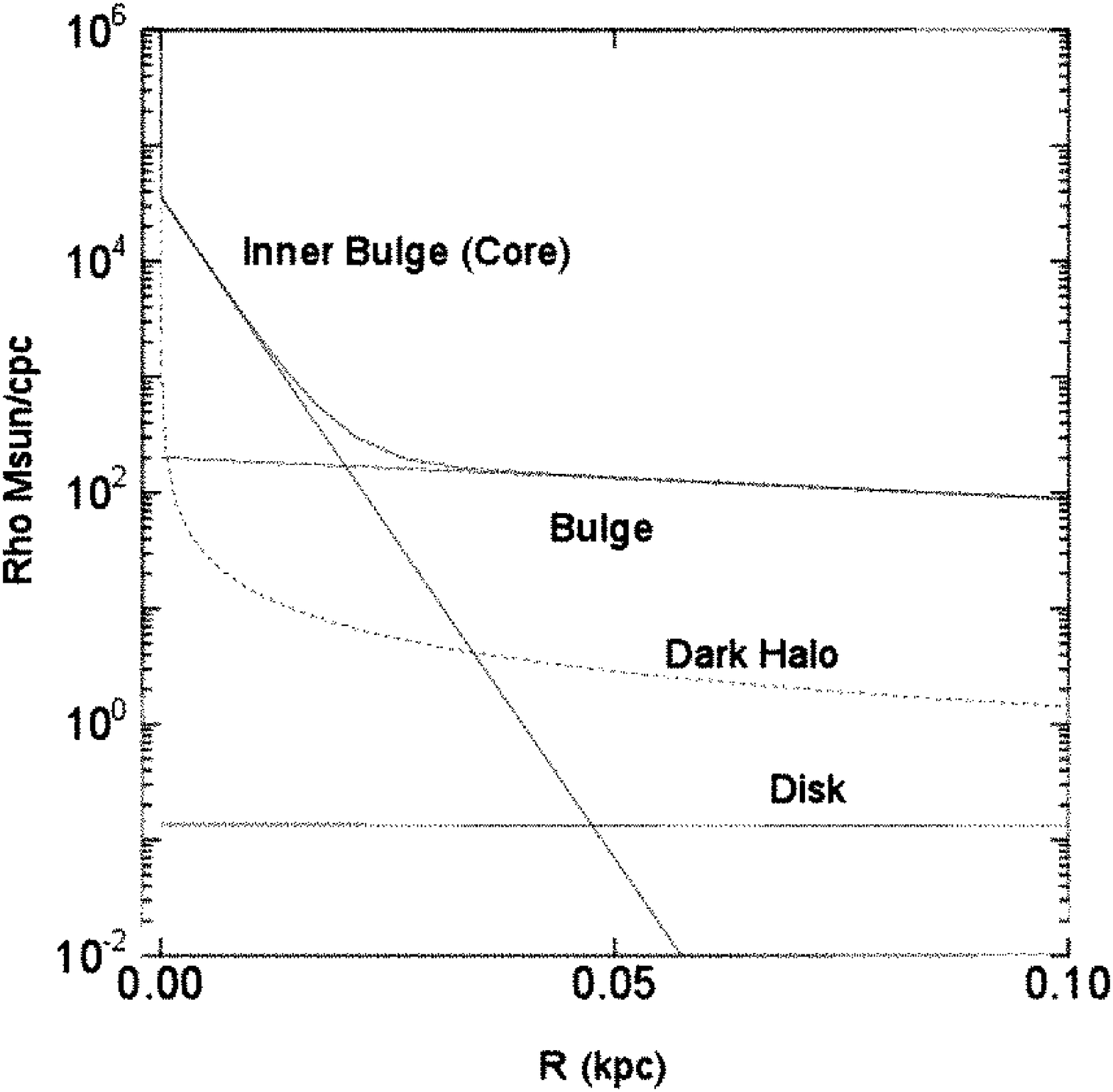}   
\ec
\caption{Left: Logarithmic plots of volume density of the exponential spheroids, exponential disk, and NFW halo calculated for the fitted parameters. Right: Same but by semi-logarithmic plot for the innermost region.}
\label{rho}
\end{figure*}

Figure \ref{rho} shows the resulting volume density profiles in the entire Galaxy for the total and individual mass components as functions of radius in logarithmic presentation. The bottom panel shows the same but for the Galactic Center in semi-logarithmic scaling. 

The calculated dark matter distribution shows a steep cusp near the nucleus because of the $1/x$ factor in the NFW profile. However, since the functional form was derived from numerical simulations with much broader resolution (Navarro et al. 1996), the exact behavior in the immediate vicinity of the nucleus may not be taken so serious, but it may include significant uncertainties.

\subsection{Surface mass density}

\begin{figure*}
\bc
\includegraphics[width=8.5cm]{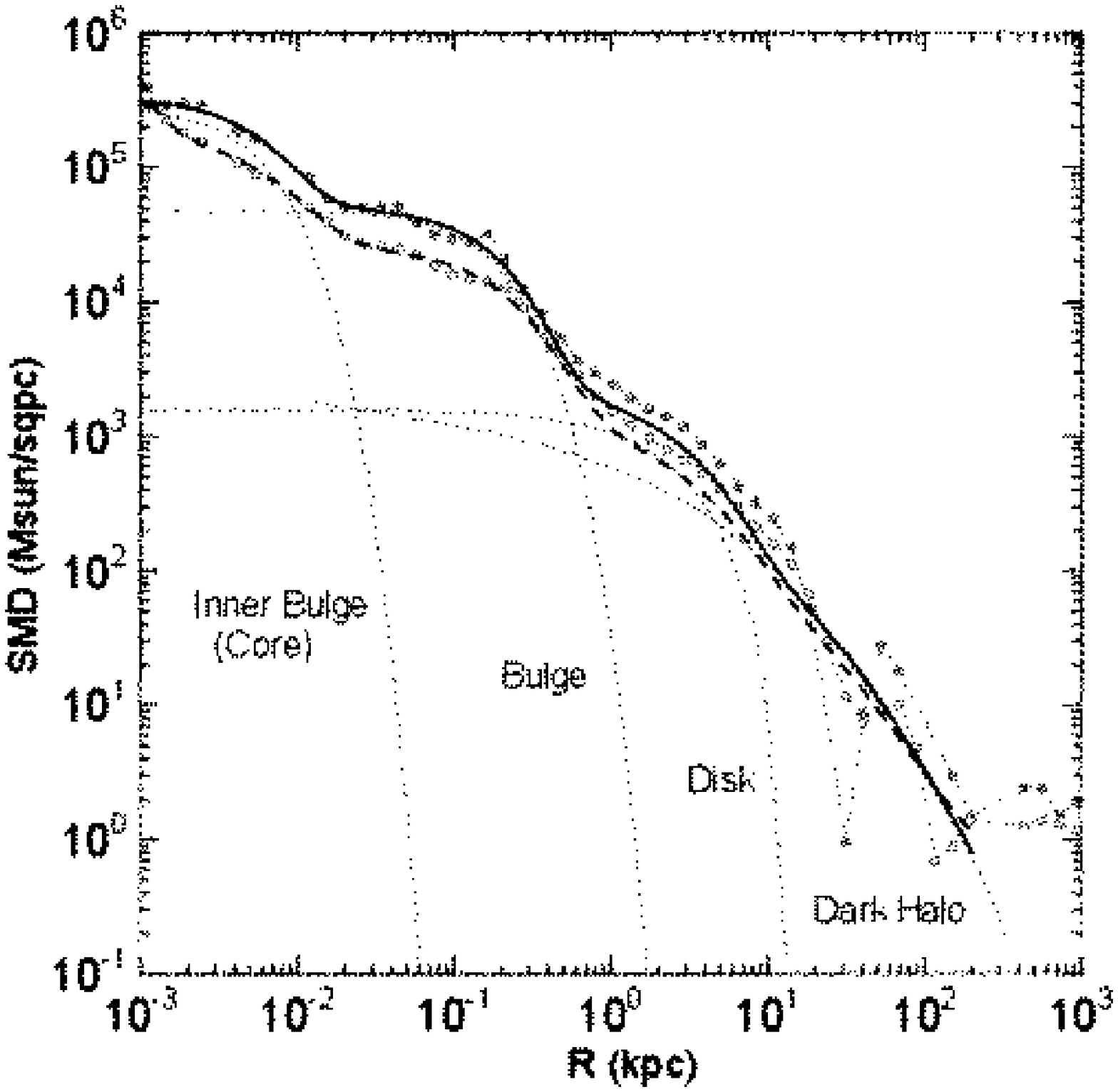}   
\includegraphics[width=8.5cm]{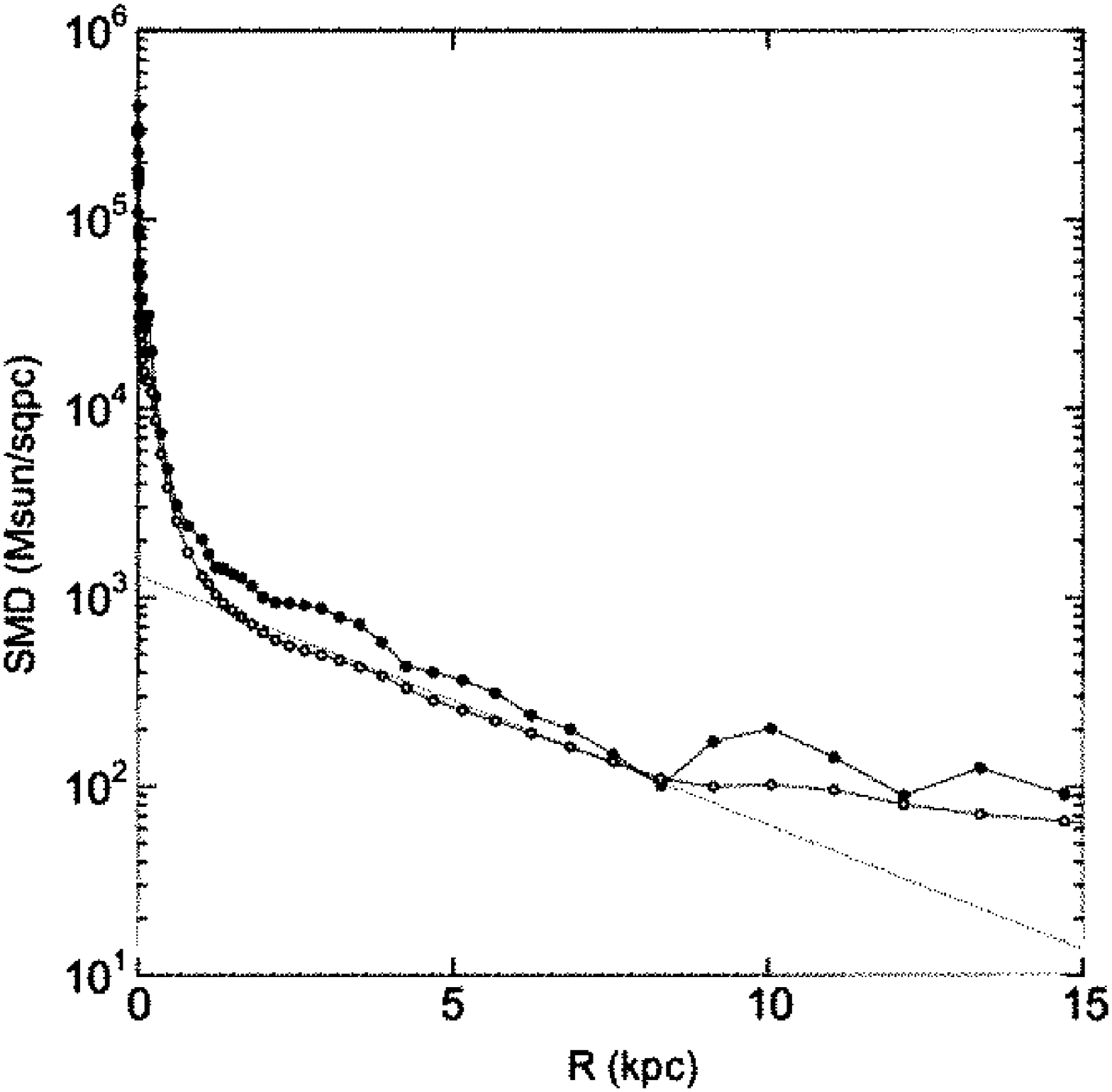}  
\ec
\caption{Left: Surface-mass density profiles. Solid line shows SMD calculated by using the model rotation curve in spherical assumption. Thin dashed lines show individual components. Thick dashed line is SMD in flat-disk assumption. Grey dots and open circles show SMD calculated by using the observed rotation curve in spherical and flat-disk assumptions, respectively. Right: Same, but direct mass alone in semi-logarithmic presentation with the exponential disk by dashed line.} 
\label{SMD}
\end{figure*}

Using the best-fit model rotation curve, we calculated the surface-mass density (SMD) as a function of radius. Figure \ref{SMD} shows the calculated results, both for the spherical and flat-disk assumptions by applying the method developed by Takamiya and Sofue (2000). In the figure, we also show the SMD distributions directly calculated using the observed rotation curve. The observed SMD is thus reproduced by the present model within errors of a factor of $\sim 1.5 - 2$ throughout the Galaxy. Here, the errors were estimated by eyes from the plots in the figure.

\subsection{Direct Calculation of Surface Mass Distribution}

Using the obtained rotation curve, we can also calculate the distribution of surface mass density (SMD) more directly (Takamiya and Sofue 2000). For a spherically symmetric model, the mass $M(r)$ inside the radius $r$ is calculated by using the rotation curve as:
\begin{equation}
M(r)=\frac{r {V(r)}^{2}}{G},
\end{equation}
where $V(r)$ is the rotation velocity at $r$. Then the SMD
${\Sigma}_{S}(R)$ at $R$ is calculated by,
\begin{eqnarray}
{\Sigma}_{\rm S}(R) & = & 2 \int\limits_0^{\infty} \rho (r) dz , \\
 & = & \frac{1}{2 \pi} \int\limits_R^{\infty} 
\frac{1}{r \sqrt{r^2-R^2}} \frac{dM(r)}{dr}dr .
\end{eqnarray}
Here, $R$, $r$ and $z$ are related by $r=\sqrt{R^2+z^2}$.  
The SMD for a thin flat-disk, ${\Sigma}_{D}(R)$, is derived by 
solving the Poisson's equation:
\begin{equation}
{\Sigma}_{\rm D}(R) =\frac{1}{{\pi}^2 G} \times
\end{equation}
$$
\left[ \frac{1}{R} \int\limits_0^R 
{\left(\frac{dV^2}{dr} \right)}_x K \left(\frac{x}{R}\right)dx + 
\int\limits_R^{\infty} {\left(\frac{dV^2}{dr} \right)}_x K \left
(\frac{R}{x}\right) \frac{dx}{x} \right],
$$
where $K$ is the complete elliptic integral and becomes very large
when $x\simeq R$ (Binney \& Tremaine 2008).   

Figure \ref{SMD} shows the calculated SMD both for spherical-symmetric mass model and for flat-disk model from $R=1$ pc to 1 Mpc. Both results agree with each other within a scatter of a factor of two. The radial profiles of SMD for the two models are similar to each other. The dashed lines represent the de Vaucouleurs law and exponential law approximately representing the bulge and disk components, respectively. The bottom panel of this figure shows the same but in semi-logarithm scaling, so that the exponential disk is represented by a straight line. It is also clearly shown that the outer SMD profile beyond $R\sim 10$ kpc is significantly displaced from the disk's profile due to the dark matter halo. Table \ref{smdtab} lists the calculated SMD values.

The present SMD plot has sufficient resolution to reveal the connection between the central black hole and bulge in the central $R\sim 1$ pc to 1 kpc region. The bulge and  black hole appears to be connected by a dense core component, which fills the gap of rotation velocity between black hole and bulge.
Figure \ref{SMD} also demonstrates smooth variation of SMD from the central black hole to the outer dark halo. The bulge at $R\sim 0.5$ kpc and exponential disk at $R\sim 3$ kpc clearly show up as the two bumps, and are followed by the dark halo extending to $\sim 400$ kpc. The mass distribution based on the grand rotation curve beyond 0.5 kpc has been extensively studied by Sofue (2012).

\section{Discussion} 

In contrast to the extensive research of the disk and outer rotation curve of the Galaxy  (e.g., Sofue and Rubin 2001; Sofue et al. 2009), the Galactic Center kinematics, particularly rotation curve and mass distribution, has not been thoroughly highlighted. This is mainly due to the too much emphasized complexity of kinematics due to the supposed bar and non-circular stream motions such as items (11) to (27) in Appendix 1.

In this paper, we abstracted simpler structures in the Galactic Center molecular line data represented by the straight LV ridges observed at high-resolution in the \co and \cs lines. We have argued that the LV ridges can represent approximate circular rotation of the dense gas disk, and obtained a central rotation curve inside $1 \sim 100$ pc. The central rotation curve was connected to the inner curve corresponding to the nuclear black hole, and also to the outer curve of the bulge, disk and dark halo. Thus, a grand rotation curve covering the entire Milky Way, from the central black hole to dark matter halo, was constructed for the first time.

\subsection{Main bulge: Exponential profile and failure in the \dv law}

The classical \dv law was found to fail to fit the observations (figure \ref{LogLogComp}).  This fact was recognized for the first time by using the logarithmic rotation curve. As argued in section 2, the \dv profile for the surface mass density requires a  central cusp, yielding steeply rising circular velocity as $V\propto r^{1/2}$ at the center. Beyond the velocity maximum at $r>a$, it declines more slowly due to the extended outskirt. Thus, the logarithmic half-maximum velocity width is about twice that for the exponential spheroid or the Plummer law as shown in figure \ref{Dfunc}. This profile was found to be inappropriate to reproduce the observations as indicated in figure \ref{LogLogComp}.

In order to reproduce the observations, we proposed a new bulge model, in which the volume density is represented by a simple exponential function as $\rho =\rho_0 e^{-r/a}$. The main bulge was found to be represented well by the exponential-spheroid model with mass $8\times 10^9 \Msun$ and scale radius 120 pc  as shown in figure \ref{LogLogComp}. 

\subsection{Inner bulge (core): Dynamical link to the black hole}

Inside the main bulge, a significant excess of rotation velocity was observed over those due to the black hole and main bulge (figure \ref{LogLogComp}). This component was well explained by an additional inner spheroidal bulge of a mass of  $\sim 5 \times 10^7 \Msun$ with the same exponential density profile as the main bulge with scale radius 3.8 pc. 

Considering the relatively large scatter and error of data at $r\sim 3-20$ pc, the density profile may not be strictly conclusive. However,  the velocity excess  should be taken as the evidence for existence of an additional mass component filling the space between the black hole and main bulge, which we called the inner bulge. As an alternative mass model to explain the plateau-like velocity excess, an isothermal sphere with flat rotation might be a candidate. However, it yields constant velocity from the center to halo, so that some artificial cut off of the sphere is required at some radius. Such a sphere with an artificial boundary may not be a good model for the Galaxy.

The Keplerian velocity by the central black hole of mass $4 \times 10^6 \Msun$ declines to 100 \kms at $r=1.5$ pc, where the observed velocities are about the same. This implies that the mass  of the inner bulge enclosed in this radius is negligible compared to the black hole mass. In fact, the present model indicates that the mass inside $r=1$ pc is only $\sim 1.2\times 10^5 \Msun$, an order of magnitude smaller than the black hole mass.

The central $\sim 1$ pc region is, therefore, controlled by the strong gravity of the massive black hole. Stars there can no longer remain as a gravitationally bound system, but are orbiting around the black hole individually by Keplerian law. As an ensemble of the stars orbiting the black hole may show velocity dispersion on the order of
$
v_{\rm \sigma}(r) \sim 125 (r/1{\rm pc})^{-1/2}{\rm km~s}^{-1}. 
$

\subsection{Comparison with the previous works}

It is worthwhile to examine if the present result is consistent with the previous works by other authors. For this purpose, we compare our result with the measurements and compilation of enclosed mass data by Genzel et al. (1994), which have been obtained using various kinds of objects such as giant stars, He I stars, HI and CO gases, circumnuclear disk, and mini spirals (See the literature therein for details). 

\begin{figure}
\bc
\includegraphics[width=8.5cm]{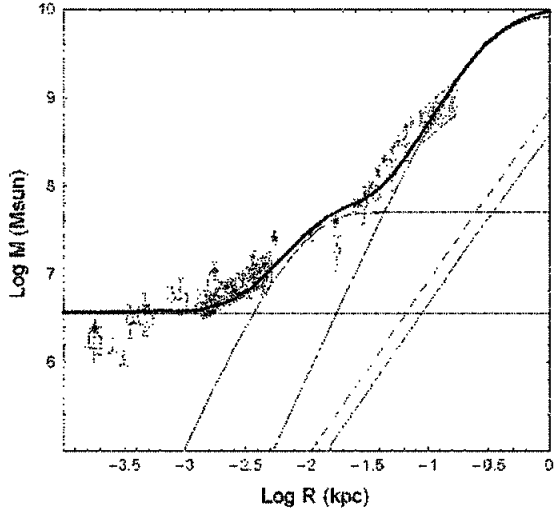}     
\ec
\caption{Comparison of enclosed mass calculated for the present rotation curve with the current results compiled by Genzel et al. (1994: see the literature therein for the plotted data). Horizontal line indicates the central black hole, thin lines show the inner bulge, main bulge and disk. The dashed line shows dark matter cusp.}
\label{mass}
\end{figure}

Figure \ref{mass} shows the enclosed mass as a function of radius calculated by the presently fitted model. In the figure we overlaid the results by Genzel et al. (1994), where their data have been converted to the case of $R_0=8.0$ kpc from 8.5 kpc adopted in their paper, multiplying the radius scale by 8.0/8.5=0.94. The mass scale was also multiplied by the same amount, as the mass is proportional to $\propto r v^2$, while radial velocities $v$ toward the Galactic Center are hardly affected by the galacto-centric distance. 

The enclosed mass for the black hole is trivially constant. The inner and main bulges have constant density near the center, which yields enclosed mass approximately proportional to $\propto r^3$ after volume integration. The disk model has constant surface density near the center, yielding enclosed mass $\propto r^2$ for surface integration.  However, it would be much less because of the finite thickness for the real galactic disk. The NFW model predicts a high density cusp near the center as $\propto r^{-1}$, yielding enclosed mass proportional to $\propto r^2$. However, the dark matter density may not be taken so serious because of the unknown accuracy of the model in the vicinity of the nucleus.

The figure shows that the present result is in good agreement with the previous observations. We stress that the wavy variation in our profile due to the two-component bulge structure is also observed in the stellar kinematics results.

\subsection{Effect of a bar and the limitation of the present analysis}

First of all, the rotation curve analysis cannot treat the non-axisymmetric part of the Galaxy (11) to (27) as listed in table \ref{parameters}. It is true that the galactic disk is superposed by non-circular streaming motions such as due to bars, arms and expanding rings. However, it is not easy to derive non-axisymmetric mass distribution from the existing observations. Simulations based on given parameters of bar potential can produce LV diagrams, and may be compared with the observations (Binney et al. 1991; Jenkins and Binney 1994; Athnasoula 1992; Burton and Liszt 1993). The present analysis would be a practical way to approach the dynamical mass structure of the central Galaxy.

We comment on the accuracy of the present analysis. The non-circular motions observed in the LV diagrams are as large as $\sim \pm 20-30$\% of the circular velocity. In the present analysis, these motions yield systematic errors of $\sim 40-60$\% of mass estimation, and the accuracy of obtained mass is about $\pm 60$\%, or within a factor of $\sim 1.6$.  The accuracy is obviously not as good as those for the outer disk and halo galactic parameters. However, it may be sufficient for examining the fundamental mass structure for the first approximation in view of the large dynamical range of order of six in the logarithmic plots of SMD as in figures \ref{rho}, \ref{SMD} and \ref{mass}.

The agreement of the present analysis with those from the stellar dynamics by Genzel et al. (1994) as shown in figure \ref{mass} may indicate that the bar's effect will not be so significant in the central several tens of parsecs. Stellar bar dynamics and stability analysis in the close vicinity of the massive black hole  would be a subject for the future. It is an interesting subject to examine if such a strong gravity by the central mass structures may allow for a long-lived bar.

\subsection{Correction for the solar rotation velocity}

Throughout the paper, we adopted the galactic parameters, $(R_0, V_0)=(8.0, 200)$ (kpc, \kms). In their recent trigonometric measurements of positions and velocities using VERA, Honma et al. (2012) obtained a faster circular velocity of the Sun of $V_0=238$ \kms. If we adopt this new value, the general rotation velocities also increases by several to $\sim 20$\% in the outer disk. We have corrected the observed rotational velocities for the difference between the new and current circular velocities of the Sun, $\Delta V=238-200=38$ \kms, using the following equation.
\be
V^{\rm c}(r)=V(r)+\Delta V \left( r \over R_0 \right).
\ee

For globular clusters and satellite galaxies, the rotation velocities were obtained by multiplying $\sqrt{2}$ to their radial velocities to  yield expected Virial velocities. We have not applied of the above correction to these cases,  because their radial velocities are influenced only statistically by the change of solar velocity, and their mean values do not significantly change by different $V_0$.

In figure \ref{LogLogComp} we plot the newly determined corrected rotation curve by open circles. The curve is not significantly changed in the central region as the above equation indicates. The outer most halo rotation curve using satellite galaxies also remains almost unchanged. A large difference is observed at $r\sim 6$ to 20 kpc, where the rotation velocity is no longer flat. The rotation velocity increases up to $\sim 20$ kpc, attaining a maximum at $V\simeq 270$ \kms. Beyond $r\sim 20$ kpc, the new rotation curve $V^{\rm c}$ declines more steeply than that for the NFW profile, but rather consistent with Keplerian curve. This indicates that the dark matter is empty beyond $\sim 20$ kpc. At this moment it is not clear if such a cut off of dark halo is indeed present, or if the simple extrapolation of the solar velocity to the outer part is allowed. 

\hskip 5mm

Acknowledgement: The author thanks  Dr. Fumi Egusa for data reduction of CO-line observations using the Nobeyama 45-m telescope, and for allowing him to use the LV diagram prior to publication of the result.

{}   

\appendix

\section{List of dynamical parameters of the Galaxy}

We list in table \ref{parameters} the dynamical parameters of the Galaxy to be determined by the rotation curve analysis. Also listed are possible parameters for the second-order perturbations causing various non-circular motions.

\begin{table*}[htbp]
\caption{Dynamical parameters for Galaxy mass study.}
\begin{center} 
\begin{tabular}{lllll}
\hline \hline

Subject
& No. & Component & Parameter & Method  \\ 

\hline
I. Axisymmetric 
& (1)&  Black hole & Mass &Stellar kinematics\\
~~~~structure
&  (2)& Bulge(s) & Mass & Rotation curve \\
&  (3)&& Radius\\
& (4)&& Profile (function)\\
& (5)&Disk & Mass\\
& (6)&& Radius \\
& (7)&& Profile (function)\\
& (8)& Dark halo& Mass\\
& (9)&& Scale radius\\
& (10)&& Profile (function)\\
\hline 
II. Non-axisymmetric
& (11)& Bar(s)& Mass&LV, $\kappa$ analysis, simulation \\
~~~~structure
& (12)&& Major axial length\\
& (13)&& Minor axial length\\
& (14)&& $z$-directional axial length\\
& (15)&& Major axis profile\\
& (16)&& Minor axis profile\\
&(17)&& z-directional profile\\
&(18)&& Position angle\\
&(19)&& Pattern speed $\Omega_{\rm p}$\\

& (20)& Arms& Density amplitude&LV, $\kappa$ analysis \\
& (21)&& Velocity amplitude\\
& (22)&& Pitch angle\\
& (23)&& Position angle\\ 
&(24)&& Pattern speed $\Omega_{\rm p}$\\
\hline 
III. Radial flow &(25)&Expanding ring&Mass&LV\\
&(26)&&Velocity\\
&(27)&&Radius\\
\hline 
\end{tabular}
\end{center}
(Note) LV stands for longitude-velocity diagram; $\kappa$ and $\Omega_{\rm p}$ are the epicyclic frequency and pattern speed. The bulge and bar may be multiple, increasing the number of parameters.
\label{parameters}
\end{table*}

\section{Tables of rotation curve and surface mass density of the Galaxy}

In this appendix, we present the observed rotation curve and surface mass density (SMD) of the entire Galaxy in tables. Table \ref{rctab} lists the radius $r$, Gaussian width of the running mean radius $\delta r$, observed rotation velocity $V$ and its statistical error $\delta V$. Table \ref{smdtab} lists SMD, $\Sigma_{\rm s}$ and $\Sigma_{\rm f}$, calculated using the rotation curve in table \ref{rctab} for spherical symmetry and thin-disk assumptions, respectively. Digitized data are available from http://www.ioa.s.u-tokyo.ac.jp/~sofue/htdocs/2013rc .

\begin{table*}[htbp]
\caption{Rotation curve of the Galaxy.}
\begin{center}
\begin{tabular}{cc}
\begin{minipage}{0.52\hsize}
\begin{center}
\begin{tabular}{llll}
\hline
$r$ (kpc) & $\delta r$ (kpc) & $V$ (\kms) & $\delta V$ (\kms) \\ 
\hline
  0.112E-02& 0.255E-03& 0.121E+03& 0.182E+02\\
  0.146E-02& 0.359E-03& 0.108E+03& 0.172E+02\\
  0.190E-02& 0.463E-03& 0.975E+02& 0.160E+02\\
  0.247E-02& 0.102E-02& 0.935E+02& 0.201E+02\\
  0.418E-02& 0.546E-03& 0.979E+02& 0.300E+02\\
  0.543E-02& 0.122E-02& 0.982E+02& 0.308E+02\\
  0.119E-01& 0.264E-02& 0.109E+03& 0.425E+02\\
  0.155E-01& 0.266E-02& 0.112E+03& 0.418E+02\\
  0.202E-01& 0.549E-02& 0.107E+03& 0.371E+02\\
  0.262E-01& 0.609E-02& 0.996E+02& 0.216E+02\\
  0.341E-01& 0.649E-02& 0.971E+02& 0.138E+02\\
  0.443E-01& 0.128E-01& 0.104E+03& 0.244E+02\\
  0.576E-01& 0.206E-01& 0.128E+03& 0.335E+02\\
  0.748E-01& 0.167E-01& 0.137E+03& 0.258E+02\\
  0.973E-01& 0.171E-01& 0.140E+03& 0.337E+02\\
  0.126E+00& 0.269E-01& 0.145E+03& 0.469E+02\\
  0.164E+00& 0.546E-01& 0.156E+03& 0.606E+02\\
  0.214E+00& 0.740E-01& 0.214E+03& 0.677E+02\\
  0.278E+00& 0.733E-01& 0.243E+03& 0.376E+02\\
  0.361E+00& 0.101E+00& 0.249E+03& 0.199E+02\\
  0.470E+00& 0.108E+00& 0.247E+03& 0.117E+02\\
  0.610E+00& 0.135E+00& 0.240E+03& 0.121E+02\\
  0.794E+00& 0.234E+00& 0.227E+03& 0.133E+02\\
  0.102E+01& 0.750E-01& 0.216E+03& 0.298E+01\\
  0.112E+01& 0.830E-01& 0.215E+03& 0.279E+01\\
  0.123E+01& 0.976E-01& 0.213E+03& 0.683E+01\\
  0.136E+01& 0.117E+00& 0.207E+03& 0.785E+01\\
  0.149E+01& 0.110E+00& 0.204E+03& 0.600E+01\\
  0.164E+01& 0.116E+00& 0.201E+03& 0.478E+01\\
  0.181E+01& 0.163E+00& 0.200E+03& 0.349E+01\\
  0.199E+01& 0.149E+00& 0.199E+03& 0.595E+01\\
  0.219E+01& 0.163E+00& 0.196E+03& 0.722E+01\\
  0.240E+01& 0.166E+00& 0.193E+03& 0.601E+01\\
  0.265E+01& 0.193E+00& 0.192E+03& 0.433E+01\\
\hline
\end{tabular}
\end{center}
\end{minipage}
\begin{minipage}{0.52\hsize}
\begin{center}
\begin{tabular}{llll}
\hline
$r$ (kpc) & $\delta r$ (kpc) & $V$ (\kms) & $\delta V$ (\kms) \\ 
\hline
  0.291E+01& 0.209E+00& 0.193E+03& 0.529E+01\\
  0.320E+01& 0.232E+00& 0.197E+03& 0.779E+01\\
  0.352E+01& 0.251E+00& 0.201E+03& 0.797E+01\\
  0.387E+01& 0.282E+00& 0.207E+03& 0.101E+02\\
  0.426E+01& 0.306E+00& 0.211E+03& 0.843E+01\\
  0.469E+01& 0.349E+00& 0.209E+03& 0.687E+01\\
  0.515E+01& 0.355E+00& 0.208E+03& 0.838E+01\\
  0.567E+01& 0.421E+00& 0.208E+03& 0.120E+02\\
  0.624E+01& 0.426E+00& 0.209E+03& 0.166E+02\\
  0.686E+01& 0.521E+00& 0.207E+03& 0.130E+02\\
  0.755E+01& 0.464E+00& 0.205E+03& 0.965E+01\\
  0.830E+01& 0.587E+00& 0.200E+03& 0.158E+02\\
  0.913E+01& 0.691E+00& 0.186E+03& 0.242E+02\\
  0.100E+02& 0.719E+00& 0.182E+03& 0.352E+02\\
  0.110E+02& 0.825E+00& 0.190E+03& 0.496E+02\\
  0.122E+02& 0.830E+00& 0.192E+03& 0.499E+02\\
  0.134E+02& 0.101E+01& 0.184E+03& 0.545E+02\\
  0.147E+02& 0.127E+01& 0.189E+03& 0.647E+02\\
  0.185E+02& 0.593E+01& 0.189E+03& 0.599E+02\\
  0.240E+02& 0.861E+01& 0.182E+03& 0.706E+02\\
  0.312E+02& 0.113E+02& 0.172E+03& 0.751E+02\\
  0.406E+02& 0.144E+02& 0.157E+03& 0.645E+02\\
  0.528E+02& 0.154E+02& 0.136E+03& 0.563E+02\\
  0.686E+02& 0.185E+02& 0.143E+03& 0.496E+02\\
  0.892E+02& 0.191E+02& 0.135E+03& 0.582E+02\\
  0.116E+03& 0.298E+02& 0.104E+03& 0.641E+02\\
  0.151E+03& 0.555E+02& 0.779E+02& 0.642E+02\\
  0.196E+03& 0.812E+02& 0.902E+02& 0.110E+03\\
  0.431E+03& 0.133E+03& 0.613E+02& 0.809E+02\\
  0.560E+03& 0.173E+03& 0.766E+02& 0.848E+02\\
  0.728E+03& 0.122E+03& 0.104E+03& 0.852E+02\\
  0.946E+03& 0.197E+03& 0.107E+03& 0.874E+02\\
  0.123E+04& 0.274E+03& 0.143E+03& 0.101E+03\\
  0.160E+04& 0.373E+03& 0.168E+03& 0.961E+02\\
\hline
\end{tabular}
\end{center}
\end{minipage}
\end{tabular}
\end{center}
\label{rctab}
\end{table*}

\begin{table*}[htbp]
\caption{Surface mass density (SMD) of the Galaxy calculated for the rotation curve in table \ref{rctab}.}
\begin{center}
\begin{tabular}{cc}
\begin{minipage}{0.45\hsize}
\begin{center}
\begin{tabular}{lll}
\hline
$r$ (kpc) & $\Sigma_{\rm s}(\Msun {\rm pc}^{-2})$ & 
$\Sigma_{\rm f}(\Msun {\rm pc}^{-2})$ \\ 
\hline
  0.112E-02& 0.287E+06& 0.395E+06\\
  0.146E-02& 0.288E+06& 0.225E+06\\
  0.190E-02& 0.302E+06& 0.177E+06\\
  0.247E-02& 0.289E+06& 0.154E+06\\
  0.418E-02& 0.183E+06& 0.109E+06\\
  0.543E-02& 0.163E+06& 0.892E+05\\
  0.119E-01& 0.824E+05& 0.499E+05\\
  0.155E-01& 0.577E+05& 0.386E+05\\
  0.202E-01& 0.499E+05& 0.303E+05\\
  0.262E-01& 0.496E+05& 0.266E+05\\
  0.341E-01& 0.502E+05& 0.252E+05\\
  0.443E-01& 0.506E+05& 0.245E+05\\
  0.576E-01& 0.380E+05& 0.219E+05\\
  0.748E-01& 0.310E+05& 0.181E+05\\
  0.973E-01& 0.285E+05& 0.156E+05\\
  0.126E+00& 0.270E+05& 0.142E+05\\
  0.164E+00& 0.311E+05& 0.139E+05\\
  0.214E+00& 0.200E+05& 0.123E+05\\
  0.278E+00& 0.115E+05& 0.867E+04\\
  0.361E+00& 0.742E+04& 0.574E+04\\
  0.470E+00& 0.481E+04& 0.382E+04\\
  0.610E+00& 0.308E+04& 0.253E+04\\
  0.794E+00& 0.239E+04& 0.173E+04\\
  0.102E+01& 0.203E+04& 0.130E+04\\
  0.112E+01& 0.169E+04& 0.117E+04\\
  0.123E+01& 0.144E+04& 0.103E+04\\
  0.136E+01& 0.142E+04& 0.927E+03\\
  0.149E+01& 0.134E+04& 0.851E+03\\
  0.164E+01& 0.127E+04& 0.785E+03\\
  0.181E+01& 0.115E+04& 0.721E+03\\
  0.199E+01& 0.999E+03& 0.654E+03\\
  0.219E+01& 0.939E+03& 0.594E+03\\
  0.240E+01& 0.933E+03& 0.553E+03\\
 \hline
\end{tabular}
\end{center}
\end{minipage}

\begin{minipage}{0.45\hsize}
\begin{center}
\begin{tabular}{lll}
\hline
$r$ (kpc) & $\Sigma_{\rm s}(\Msun {\rm pc}^{-2})$ & 
$\Sigma_{\rm f}(\Msun {\rm pc}^{-2})$ \\ 
\hline
  0.265E+01& 0.909E+03& 0.523E+03\\
  0.291E+01& 0.877E+03& 0.496E+03\\
  0.320E+01& 0.786E+03& 0.464E+03\\
  0.352E+01& 0.721E+03& 0.428E+03\\
  0.387E+01& 0.577E+03& 0.385E+03\\
  0.426E+01& 0.431E+03& 0.330E+03\\
  0.469E+01& 0.402E+03& 0.285E+03\\
  0.515E+01& 0.366E+03& 0.252E+03\\
  0.567E+01& 0.313E+03& 0.223E+03\\
  0.624E+01& 0.239E+03& 0.191E+03\\
  0.686E+01& 0.200E+03& 0.161E+03\\
  0.755E+01& 0.148E+03& 0.135E+03\\
  0.830E+01& 0.101E+03& 0.110E+03\\
  0.913E+01& 0.173E+03& 0.999E+02\\
  0.100E+02& 0.203E+03& 0.102E+03\\
  0.110E+02& 0.142E+03& 0.959E+02\\
  0.122E+02& 0.895E+02& 0.803E+02\\
  0.134E+02& 0.125E+03& 0.714E+02\\
  0.147E+02& 0.908E+02& 0.657E+02\\
  0.185E+02& 0.508E+02& 0.449E+02\\
  0.240E+02& 0.284E+02& 0.273E+02\\
  0.312E+02& 0.137E+02& 0.161E+02\\
  0.406E+02& 0.646E+01& 0.915E+01\\
  0.528E+02& 0.151E+02& 0.704E+01\\
  0.686E+02& 0.412E+01& 0.545E+01\\
  0.116E+03& 0.661E+00& 0.995E+00\\
  0.151E+03& 0.428E+01& 0.156E+01\\
  0.196E+03& 0.174E+01& 0.176E+01\\
  0.431E+03& 0.242E+01& 0.133E+01\\
  0.560E+03& 0.241E+01& 0.143E+01\\
  0.728E+03& 0.154E+01& 0.130E+01\\
  0.946E+03& 0.189E+01& 0.119E+01\\
  0.123E+04& 0.114E+01& 0.108E+01\\
\hline
\end{tabular}
\end{center}
\end{minipage}
\end{tabular}
\end{center}
\label{smdtab}
\end{table*}


\begin{thebibliography}{}
\def\r{\bibitem[]{}} 


\r Athanassoula, E. 1992 MNRAS 259, 345. 

\bibitem[Bally et al.(1987)]{1987ApJS...65...13B} Bally, J., Stark, A.~A., Wilson, R.~W., \& Henkel, C.\ 1987, \apjs, 65, 13 

\r Binney, J. and Tremain, S. 2008, in "Galactic Dynamics", 2nd ed., Chap. 2.

\bibitem[Binney et al.(1991)]{1991MNRAS.252..210B} Binney, J., Gerhard, O.~E., Stark, A.~A., Bally, J., \& Uchida, K.~I.\ 1991, \mnras, 252, 210  

\r Burton W B, Gordon M A. 1978. {AA} 63: 

\r Burton, W. B. and Liszt H. S. 1993 AA 274, 765.

\r Clemens, D. P. 1985. {\it Ap. J.} 295:422     

\bibitem[Crawford et al.(1985)]{1985Natur.315..467C} Crawford, M.~K., 
Genzel, R., Harris, A.~I., et al.\ 1985, \nat, 315, 467 

\bibitem[Genzel 
\& Townes(1987)]{1987ARA&A..25..377G} Genzel, R., \& Townes, C.~H.\ 1987, \araa, 25, 377 

\bibitem[Genzel et al.(1994)]{1994RPPh...57..417G} Genzel, R., Hollenbach, D., \& Townes, C.~H.\ 1994, Reports on Progress in Physics, 57, 417 

\bibitem[Genzel et al.(2000)]{2000MNRAS.317..348G} Genzel, R., Pichon, C., Eckart, A., Gerhard, O.~E., \& Ott, T.\ 2000, \mnras, 317, 348 

\bibitem[Genzel et al.(2010)]{2010RvMP...82.3121G} Genzel, R., Eisenhauer, F., \& Gillessen, S.\ 2010, Reviews of Modern Physics, 82, 3121 


\bibitem[Ghez et al.(2005)]{2005ApJ...620..744G} Ghez, A.~M., Salim, S., Hornstein, S.~D., et al.\ 2005, \apj, 620, 744 

\bibitem[Ghez et al.(2008)]{2008ApJ...689.1044G} Ghez, A.~M., Salim, S., 
Weinberg, N.~N., et al.\ 2008, \apj, 689, 1044 

\bibitem[Gillessen et al.(2009)]{2009ApJ...692.1075G} Gillessen, S., Eisenhauer, F., Trippe, S., et al.\ 2009, \apj, 692, 1075 

\bibitem[Honma et al.(2012)]{2012PASJ...64..136H} Honma, M., Nagayama, T., Ando, K., et al.\ 2012, \pasj, 64, 136  

\bibitem[Jenkins \& Binney(1994)]{1994MNRAS.270..703J} Jenkins, A., \& Binney, J.\ 1994, \mnras, 270, 703  

\r Kaifu, N. Kato, T., Iguchi, T. 1972 Nature, 238, 105.

\bibitem[Lindqvist et al.(1992)]{1992A&A...259..118L} Lindqvist, M., Habing, H.~J., \& Winnberg, A.\ 1992, \aap, 259, 118 

\r Navarro, J. F., Frenk, C. S., White, S. D. M., 1996, ApJ, 462, 563   

\bibitem[Oh et al.(2010)]{2010PASJ...62..101O} Oh, C.~S., Kobayashi, H., Honma, M., et al.\ 2010, \pasj, 62, 101 


\bibitem[Oka et al.(1998)]{1998ApJS..118..455O} Oka, T., Hasegawa, T., Sato, F., Tsuboi, M., \& Miyazaki, A.\ 1998, \apjs, 118, 455 

\bibitem[Rieke \& Rieke(1988)]{1988ApJ...330L..33R} Rieke, G.~H., \& Rieke, M.~J.\ 1988, \apjl, 330, L33 

\bibitem[Sakai et al.(2012)]{2012PASJ...64..108S} Sakai, N., Honma, M., Nakanishi, H., et al.\ 2012, \pasj, 64, 108 

\r Scoville, N. 1972, ApJ 175, L127.

\bibitem[Sofue(2012)]{2012PASJ...64...75S} Sofue, Y.\ 2012, \pasj, 64, 75  

\bibitem[Sofue et al.(2009)]{2009PASJ...61..227S} Sofue, Y., Honma, M., \& Omodaka, T.\ 2009, \pasj, 61, 227 

\r Sofue, Y., Rubin, V.C. 2001 ARAA 39, 137  

\bibitem[Sofue(1995)]{1995PASJ...47..551S} Sofue, Y.\ 1995, \pasj, 47, 551 

\bibitem[Sofue(1995)]{1995PASJ...47..527S} Sofue, Y.\ 1995, \pasj, 47, 527 

\bibitem[Sofue(1996)]{1996ApJ...458..120S} Sofue, Y.\ 1996, \apj, 458, 120 

\bibitem[Sofue(2009)]{2009PASJ...61..153S} Sofue, Y.\ 2009, \pasj, 61, 153 

\bibitem[Sofue(2012)]{2012PASJ...64...75S} Sofue, Y.\ 2012, \pasj, 64, 75  

\bibitem[Sofue et al.(2009)]{2009PASJ...61..227S} Sofue, Y., Honma, M., \& Omodaka, T.\ 2009, \pasj, 61, 227 

\bibitem[Takamiya \& Sofue(2000)]{2000ApJ...534..670T} Takamiya, T., \& Sofue, Y.\ 2000, \apj, 534, 670 

\bibitem[Tsuboi et al.(1999)]{1999ApJS..120....1T} Tsuboi, M., Handa, T., \& Ukita, N.\ 1999, \apjs, 120, 1 

\bibitem[Xue et al.(2008)]{2008ApJ...684.1143X} Xue, X.~X., Rix, H.~W., 
Zhao, G., et al.\ 2008, \apj, 684, 1143 


\end{thebibliography}
\end{document}